\documentclass{jpp}
\usepackage{mydefs}

\shorttitle{Mean-field dynamo as a quantumlike modulational instability}
\shortauthor{S. Jin and I. Y. Dodin}

\title{Mean-field dynamo as a quantumlike modulational instability}

\author{S. Jin\aff{1, 2}
  \corresp{\email{sjin@pppl.gov}} \and
  I. Y. Dodin\aff{1, 2}}

\affiliation{\aff{1}Department of Astrophysical Sciences, Princeton University, Princeton, New Jersey 08543, USA
\aff{2}Princeton Plasma Physics Laboratory, Princeton, New Jersey 08540, USA}

\begin{document}

\maketitle

\begin{abstract}
Presented here is a novel formulation of the mean-field dynamo as a modulational instability of magnetohydrodynamic (MHD) turbulence. This formulation, termed mean-field wave kinetics (MFWK), is based on the Weyl symbol calculus and allows describing the interaction between the mean fields (magnetic field and fluid velocity) and turbulence without requiring scale separation that is commonly assumed in literature. The turbulence is described by the Wigner--Moyal equation for the spectrum of the two-point correlation matrix (Wigner matrix) of magnetic-field and velocity fluctuations and depicts the turbulence as an effective plasma of quantumlike particles that interact via the mean fields. Eddy--eddy interactions, which serve as `collisions' in this effective plasma, are modeled within the standard minimal tau approximation to aid comparison with existing theories. Using MFWK, the nonlocal electromotive force is calculated for generic turbulence from first principles, modulo the limitations of MFWK. This result is then used to study, both analytically and numerically, the modulational modes of MHD turbulence, which appear as linear instabilities of said effective quantumlike plasma of fluctuations. The standard $\alpha^2$-dynamo and other known results are reproduced as a special cases. A new dynamo effect is predicted that is driven by correlations between the turbulent flow velocity and the turbulent current.
\end{abstract}

\section{Introduction}
\label{sec:intro}

\subsection{Background}
The universe abounds with turbulent magnetised plasma \citep{ref:schekochihin07}. It is now widely accepted that the observed (or inferred) strength of the magnetic fields that thread these astrophysical systems cannot be explained without invoking some kind of turbulent-dynamo process, in which the kinetic energy of the turbulent plasma flows is converted into magnetic energy \citep{ref:vainshtein72,ref:brandenburg05,ref:subramanian19,ref:federrath16}. These magnetic fields are often correlated on scales large compared to those of the underlying turbulence, and it is well established that the dynamics of these orderly fields are inextricably linked with the underlying plasma turbulence \citep{ref:tobias21, ref:brandenburg23}.\footnote{A classic example is our own sun, which exhibits a large-scale dipolar magnetic field that reverses polarity on a curious 11-year cycle \citep{ref:charbonneau14,ref:jones10}.} Still,
many aspects of this so-called `large-scale' turbulent dynamo remain to be understood \citep{ref:radler14,ref:hughes18}. 

One of the mainstream tools for studying the large-scale turbulent dynamo has been mean-field electrodynamics \citep{book:krause80,book:moffatt78,ref:roberts75,ref:shukurov21}. In this approach, the magnetohydrodynamic (MHD) model of interest is decomposed into a system of coupled equations for both the turbulent and mean-field components, and, ultimately, a closed equation is obtained that describes the evolution of the mean magnetic field in response to correlations of the turbulent fluctuations. Mean-field electrodynamics has now profoundly shaped much of our understanding of cosmical dynamos \citep{ref:hughes18}, and there have been many refinements to this basic framework over the years \citep{ref:radler14,ref:brandenburg18}. However, two key aspects of the turbulent dynamo continue to pose substantial challenges for the mean-field approach.

The first of these challenges arises from the so-called small-scale dynamo. Historically, the mean-field dynamo problem was one of coherent magnetic fields emerging from a state of hydrodynamic turbulence. This original understanding of the problem justified the kinematic approximation, in which one takes the flows to be prescribed, as the magnetic fields of interest would be too weak to substantially modify these flows, at least in the initial linear growth stage. Within this approximation, one need not consider the full MHD system and can instead focus on the much simpler task of solving the induction equation for a given flow. However, it is now understood that in the astrophysically interesting regime of large magnetic Reynolds number (Rm), the small-scale dynamo generates substantial turbulent magnetic fields on time scales much shorter than those of the large-scale dynamo modes predicted by kinematic mean-field theory \citep{ref:rincon19}. 
It has therefore been suggested that a better-posed mean-field problem is one in which mean fields grow from the MHD turbulence aftermath of a saturated small-scale dynamo \citep{ref:rincon19,ref:tobias21}. As will be discussed extensively throughout this paper, for MHD turbulence, the mean magnetic and velocity fields can be dynamically coupled, such that the solution of the full MHD mean-field problem requires a self-consistent mean-field treatment for both the velocity and magnetic fields. Although the importance of such an approach has been recognised \citep{ref:tobias21,ref:rincon19}, and some progress has been made--analytically for simple cases as in \citep{ref:courvoisier10,ref:courvoisier10b}, and numerically with direct statistical simulations (DSS) \citep{ref:mondal23}--the basic theory of mean-field modes for generic MHD turbulence has been lacking.

Another significant limitation of existing mean-field theories is that most of them rely on the assumption that mean fields vary on much longer scales than the turbulent fields. This assumed separation of scales is used to simplify the problem by enabling a local closure in which the turbulent EMF at a given point in spacetime is expressed only in terms of the mean fields and their low-order derivatives at that same point in spacetime.
However, the assumption of scale separation is often poorly justified, and there has been a growing interest in understanding the impact of so-called nonlocal effects on mean-field dynamics \citep{ref:pipin23,ref:bendre22,ref:rheinhardt14,ref:rheinhardt12,ref:hubbard09,ref:brandenburg08,ref:gressel20}. But thus far, these effects have been studied mostly numerically, and, although some analytical calculations do exist \citep{ref:rudiger01}, they have been largely intractable for generic turbulence.
\subsection{Mean-field wave kinetics\label{sec:mfwkintro}}
In this paper, we propose a novel formalism, mean-field wave kinetics (MFWK), that is able to overcome the aforementioned limitations. MFWK retains the dynamical independence of the turbulent correlations and treats them within the wave-kinetics framework. 

To introduce the general idea of wave kinetics, let us first consider a simplified problem where the characteristic wavelength of the turbulent fluctuations is much less than the inhomogeneity scale, 
\begin{equation}\label{eq:gocond}
    l/L \ll 1.
\end{equation} In this `geometrical-optics' (GO) limit, waves are much like particles in the sense that they can be described by Hamilton's ray equations \citep{book:tracy14,my:ql}:
\begin{equation}
    \dot{\vec{x}}=\pd_{\vec{k}}\omega, \qquad \dot{\vec{k}}=-\pd_{\vec{x}}\omega.
\end{equation}
Here, \m{\vec{x}} is the ray position, \m{\vec{k}} is the local wavevector (which is proportional to the ray's canonical momentum), and \m{\omega} is the local frequency and serves as the ray Hamiltonian.
Wave kinetics therefore refers to the kinetic theory that describes the phase-space dynamics of these quasiparticles.

Turbulence can be viewed as an ensemble of such quasiparticles, with a distribution function in phase space evolving due to wave--wave collisions (`eddy--eddy' interactions) and collective effects (mean fields):
\begin{equation}\label{eq:kinetic}
    \pd_t f=\mc{C}[f]+\{H,f\}.
\end{equation}
The situation when the collision term \m{\mc{C}[f]} dominates over \m{\{H,f\}} corresponds to homogeneous turbulence, where the classic subject of interest is turbulent spectra \citep{ref:vedenov67,book:nazarenko11}. Here, though, we focus on the opposite limit, when \m{\mc{C}[f]} is negligible compared to the collective interactions determined by \m{\{H,f\}}. This corresponds to the regime where turbulence effectively acts as a collisionless plasma. In this sense, our formulation can be viewed as `plasma physics of turbulence' \citep{ref:tsiolis20}, and the resulting mean-field theory can be understood as a theory of collective effects in MHD turbulence. Although this approach has been widely used in the past (for an overview and references, see, for example \citet{ref:zhu21, thesis:ruiz17,book:mendonca00}), its application to MHD turbulence has been limited on the account of being technically challenging.

In particular, note that the GO approximation is typically \textit{not} satisfied when the mean-field scales themselves are formed through modulational instabilities (MIs) \citep{ref:zhu21,ref:tsiolis20}. This means that \Eq{eq:kinetic} must be replaced with a more general model that does not rely on scale separation. Furthermore, MHD fluctuations are inherently electromagnetic. This means that a proper quantum analogy for them is \textit{vector} particles (particles with spin) and thus their distribution is a matrix rather than a scalar. 
The corresponding generalisation of \Eq{eq:kinetic} can be constructed using the Weyl symbol calculus (\App{app:wignerweyl}), or more specifically, the Wigner--Moyal formalism \citep{book:weyl50}, which will be explained in detail in \Sec{sec:mfwkderiv}. This formalism provides access to regimes outside the traditional domain of the mean-field theories, yet remains analytically tractable. It depicts turbulence self-organization as a collective instability of an effective quantumlike plasma of turbulent fluctuations, in which mean fields serve as a collective field through which the fluctuations interact. The Wigner--Moyal formalism is also advantageous in that it maintains a clear connection with the GO model \Eq{eq:kinetic}, namely, subsumes it as a limit.

\subsection{Outline}
This paper aims to systematically develop MFWK for MHD turbulence (sections \ref{sec:mfwkderiv} and \ref{sec:linmod}) and use this theory to study basic physics of dynamo by considering several illustrative examples (sections \ref{sec:nonlocalemf} and \ref{sec:modmodes}). Rather than modeling a specific astrophysical system in detail (a pursuit that does not lack participants), we focus on developing the theoretical framework and exploring its broader implications. We find that even for the simple examples considered here, MFWK predicts qualitatively distinct features compared to previous mean-field theories. 

In our first application of MFWK, we derive, from first principles, the nonlocal-response kernel of the turbulent EMF for weak mean fields and generic MHD turbulence. We then evaluate this expression for the special cases of (i) hydrodynamic and (ii) isotropic MHD turbulence. For hydrodynamic turbulence, we find that MFWK theoretically predicts the same generic form of the nonlocal EMF that has been commonly assumed in the literature based on simulations \citep{ref:brandenburg23}. For isotropic MHD turbulence, the scale-separated limit of MFWK largely reproduces the predictions of existing mean-field theories with the important exception of the dependence of the EMF on the mean flow. However, we report a qualitatively different dependence of the EMF on the mean flow compared to the only other (to our knowledge) existing calculation of this effect \citep{ref:radler10}.

In the second application of MFWK, we identify the mean-field effects associated with various statistical properties of homogeneous isotropic MHD turbulence. To do this, we first derive the general dispersion relation of modulational modes of generic MHD turbulence. We then solve the dispersion relation for the specific case of ideal isotropic MHD turbulence. Beyond the well-known \m{\alpha^2}-dynamo driven by kinetic helicity, we predict a new dynamo effect that is driven by correlations between the fluctuating flow and current, \m{\flav{\fl{\vec{v}}\cdot\fl{\vec{j}}}}. We also predict sound-like `correlation waves' that propagate (and, depending on the properties of turbulence, possibly grow) through plasma at speeds determined by the statistical properties of the turbulent fluctuations.

This paper is organised as follows. In \Sec{sec:mfwkderiv}, we derive the main equations of MFWK. In \Sec{sec:linmod}, we linearise the MFWK equations around generic turbulent equilibria and formulate mean-field effects in terms of modulational (in)stability. In \Sec{sec:nonlocalemf}, we apply the MFWK to derive the nonlocal turbulent EMF and analyse its properties. In \Sec{sec:modmodes}, we derive and solve the dispersion relation of modulational modes for MHD turbulence, and we also characterise the \m{\flav{\fl{\vec{v}}\cdot\fl{\vec{j}}}}-dynamo. The main results are summarised in \Sec{sec:summary}.

\section{Derivation of mean-field wave-kinetics\label{sec:mfwkderiv}}
\subsection{Base model: incompressible resistive MHD}
As a base model, we assume incompressible resistive MHD with homogeneous mass density \m{\rho=\text{const}}:

\begin{subequations}\label{eq:mhd}
\begin{gather}
 \pd_t \vec{v} + (\vec{v} \cdot \nabla) \vec{v} 
  = (\vec{b} \cdot \nabla)\vec{b} - \nabla P 
    + \nu \del^2 \vec{v},
    \label{eq:v}\\
 \pd_t \vec{b} = \nabla \times (\vec{v} \times \vec{b}) + \eta \nabla^2 \vec{b},
    \label{eq:b}\\
     \nabla \cdot \vec{v} = 0,
 \qquad
 \nabla \cdot \vec{b} = 0.
\end{gather}
\end{subequations}
Here \m{\vec{v}} is the fluid velocity; \m{\vec{b} \doteq \vec{B}/\sqrt{4\pi\rho}} is the local Alfv\'en velocity; \m{P \doteq (P_{\text{kin}} + B^2/8\pi)/\rho} is the normalised total pressure, with \m{P_{\text{kin}}} being the kinetic pressure (the symbol \m{\doteq} denotes definitions); \m{\nu} is the viscosity and \m{\eta} is the resistivity, both of which are assumed constant. In order to put \Eqs{eq:mhd} in a more symmetric form, let us rewrite them in terms of the two Els\"asser fields \m{\vec{z}^\pm} \citep{ref:elsasser50}, which are also solenoidal:
\begin{gather}\label{eq:zdef}
 \vec{z}^\pm \doteq \vec{v} \pm \vec{b},
 \qquad
 \del \cdot \vec{z}^\pm  = 0.
\end{gather}
This leads to two coupled equations for \m{\vec{z}^\pm}:
\begin{gather}\label{eq:elsasser}
\pd_t \vec{z}^\pm=-(\vec{z}^\mp\cdot\nabla)\vec{z}^\pm-\nabla P
+\nu_+ \nabla^2 \vec{z}^\pm+\nu_- \nabla^2 \vec{z}^\mp,
\end{gather}
where \m{\nu_+ \doteq (\nu + \eta)/2} and \m{\nu_- \doteq (\nu - \eta)/2}. The normalised total pressure \m{P} can be found as follows. By taking the divergence of \Eq{eq:v} and using \Eq{eq:zdef}, one obtains
\begin{gather}\label{eq:peq}
 \nabla^2 P = -\nabla\cdot[(\vec{z}^\mp \cdot \nabla)\vec{z}^\pm].
\end{gather}
Let us introduce the wavevector operator \m{\boper{k} \doteq - \ii \nabla}, so \m{\oper{k}^2 \doteq \smash{\boper{k}}^2 = - \nabla^2}. Let us also assume some appropriate (say, periodic) boundary conditions. Then, \Eq{eq:peq} yields
\begin{gather}\label{eq:pinvert}
 P = -\oper{k}^{-2}\boper{k}\cdot[(\vec{z}^\mp \cdot \boper{k})\vec{z}^\pm] + \const,
\end{gather}
whence \m{\nabla P} in \Eq{eq:elsasser} can be expressed through \m{\vec{z}^\pm}. 

Alternatively, \m{\nabla P} can be eliminated from \Eq{eq:elsasser} by taking the curl of this equation:
\begin{gather}\label{eq:elsasservort}
\pd_t \vec{w}^\pm=-\nabla\times[(\vec{z}^\mp\cdot\nabla)\vec{z}^\pm]
+\nu_+ \nabla^2 \vec{w}^\pm+\nu_- \nabla^2 \vec{w}^\mp,
\end{gather}
where the Els\"asser vorticities \m{\vec{w}^\pm } are defined:
\begin{gather}\label{eq:wintermsofz}
\vec{w}^\pm \doteq \nabla \times \vec{z}^\pm.
\end{gather}
Indeed, due to \Eq{eq:zdef}, one can express \m{\vec{z}^\pm} using a vector potential \m{\vec{a}^\pm} such that \m{\vec{z}^\pm = \nabla \times \vec{a}^\pm}. Let us assume the gauge such that \m{\nabla \cdot \vec{a}^\pm = 0}. Then,
\begin{gather}
 \vec{w}^\pm = \nabla \times (\nabla \times \vec{a}^\pm)
 = - \nabla^2 \vec{a}^\pm \equiv \oper{k}^2 \vec{a}^\pm,
\end{gather}
whence
\begin{gather}\label{eq:zintermsofw}
 \vec{z}^\pm = \ii \oper{k}^{-2} (\boper{k} \times \vec{w}^\pm).
\end{gather}
(Here, we assume that \m{\vec{z}^\pm} have zero spatial average, so \m{\oper{k}^2} is invertible.) Then, \Eq{eq:elsasser} can be expressed through \m{\vec{w}^\pm} alone:
\begin{gather}
\pd_t \vec{w}^\pm = 
-\boper{k} \times
\{
[(\boper{k} \times \oper{k}^{-2}\vec{w}^\mp)\cdot \boper{k}]
(\boper{k}\times\oper{k}^{-2}\vec{w}^\pm)
\}
-\oper{k}^2(\nu_+ \vec{w}^\pm + \nu_- \vec{w}^\mp).
\label{eq:vecw}
\end{gather}

Writing the Elsässer equations in this form has the benefit that the nonlinear term is now expressed as a product of inverse operators that act only on the Els\"sser vorticities themselves, rather than an inverse operator that acts on the product of the Elsässer fields as in \Eq{eq:pinvert}.
As we will see shortly, this property makes \Eq{eq:vecw} more convenient than \Eq{eq:elsasser} for the
eventual formulation of MFWK. Note also that \Eq{eq:vecw} is equivalent to \Eq{eq:elsasservort} with the assumption that \m{\vec{z}^\pm} have zero spatial average.

\subsection{Mean fields vs. fluctuations}
As in the usual mean-field approach \citep{book:krause80,book:moffatt78}, we decompose our Els\"asser fields into mean and fluctuating parts:
\begin{equation}\label{eq:meanfluc}
    \vec{z}^\pm=\av{\vec{z}}^\pm+\fl{\vec{z}}^\pm, \qquad \vec{w}^\pm=\av{\vec{w}}^\pm+\fl{\vec{w}}^\pm.
\end{equation}
Inserting \Eq{eq:meanfluc} into \Eq{eq:elsasservort} and averaging yields the following equation for the mean fields:
\begin{equation}\label{eq:mean}
\begin{aligned}
        \pd_t \av{\vec{w}}^\pm =& -\nabla\times[(\av{\vec{z}}^\mp\cdot\nabla)\av{\vec{z}}^\pm]
+\nu_+ \nabla^2 \av{\vec{w}}^\pm+\nu_- \nabla^2 \av{\vec{w}}^\mp-\flav{\nabla\times[(\fl{\vec{z}}^\mp\cdot\nabla)\fl{\vec{z}}^\pm]}.
\end{aligned}
\end{equation}
(Note that this is similar to \Eq{eq:elsasservort}, but has an additional source term due to the fluctuating fields.) Subtracting \Eq{eq:mean} from \Eq{eq:elsasservort} yields the following equation for the fluctuations:
\begin{equation}\label{eq:fluc}
\begin{aligned}
        \pd_t \fl{\vec{w}}^\pm =& -\nabla\times[(\av{\vec{z}}^\mp\cdot\nabla)\fl{\vec{z}}^\pm+(\fl{\vec{z}}^\mp\cdot\nabla)\av{\vec{z}}^\pm]
+\nu_+ \nabla^2 \fl{\vec{w}}^\pm+\nu_- \nabla^2 \fl{\vec{w}}^\mp+\vec{F}^\pm_{\text{NL}},
\end{aligned}
\end{equation}
where the nonlinear term
\begin{equation}
\vec{F}^\pm_{\text{NL}}\doteq\flav{\nabla\times[(\fl{\vec{z}}^\mp\cdot\nabla)\fl{\vec{z}}^\pm]}-\nabla\times[(\fl{\vec{z}}^\mp\cdot\nabla)\fl{\vec{z}}^\pm]
\end{equation}
corresponds to eddy--eddy interactions. 

Within the quasilinear approximation (QLA) (\ie the collisionless-wave approximation of \Sec{sec:mfwkintro}), \m{\vec{F}^\pm_{\text{NL}}} is neglected. Although nonlinear effects are, of course, important for understanding the full picture, our present focus is on understanding mean-field formation, many aspects of which can be studied within the QLA \citep{ref:zhu21,ref:tsiolis20}. As discussed in the previous section, this can also be understood as retaining collective effects while neglecting pair-wise interactions of the turbulent fluctuations. Deviations from the QLA are discussed extensively in \cite{ref:jin24}.

\subsection{Wigner--Moyal equation for the fluctuations}
\subsubsection{Basic notation\label{sec:basicnot}}

Let us first introduce the state ket vectors:
\begin{equation}
\ket{\fl{\vec{z}}}\doteq\begin{pmatrix}\ket{\fl{\vec{z}}^+}\\\ket{\fl{\vec{z}}^-}
    \end{pmatrix},\qquad \ket{\fl{\vec{w}}}\doteq\begin{pmatrix}\ket{\fl{\vec{w}}^+}\\\ket{\fl{\vec{w}}^-}
    \end{pmatrix}.
\end{equation}
The usual fields over space-time can be understood as the spatial projections of these kets, \ie \m{\braket{\vec{x}|\fl{\vec{z}}^\pm}=\fl{\vec{z}}^\pm (\vec{x})}. The fluctuation equation can then be written as a vector Schr\"odinger equation for \m{\ket{\fl{\vec{w}}}}:
\begin{equation}
    \begin{gathered}\label{eq:schrodinger}
        i\partial_t\ket{\fl{\vec{w}}}=\op{\matr{H}}' \ket{\fl{\vec{w}}},\\
\end{gathered}
\end{equation}
with the (generally non-Hermitian) Hamiltonian:
\begin{equation}\label{eq:fluchamiltonian}
        \op{\matr{H}}'\doteq\begin{pmatrix}
  \op{\matr{H}}'^{++}&\op{\matr{H}}'^{+-}\\ \op{\matr{H}}'^{-+}&\op{\matr{H}}'^{--}
\end{pmatrix},
\end{equation}
where we have introduced
\begin{subequations}
    \begin{gather}
\op{H}_{ij}^{'\pm\pm}=\delta_{ij}\Big(\av{z}_l^\mp\op{k}_l-i\av{z}_{l,m}^\mp\frac{\op{k}_l\op{k}_m}{\op{k}^2}-i\nu_+\op{k}^2\Big)+i\av{z}_{l,m}^\mp\frac{\op{k}_l\op{k}_m}{\op{k}^2},\\
\begin{aligned}
\op{H}_{ij}^{'\pm\mp}&=\delta_{ij}\Big(i\av{z}_{l,m}^\pm\frac{\op{k}_l\op{k}_m}{\op{k}^2}+\av{z}^\pm_{l,mm}\frac{\op{k}_l}{\op{k}^2}-i\nu_-\op{k}^2\Big)\\
    &\quad+i\Big(\av{z}^\pm_{j,i}-\av{z}^\pm_{j,l}\frac{\op{k}_i\op{k}_l}{\op{k}^2}\Big)+(\av{z}_{j,il}^\pm-\av{z}^\pm_{l,ij})\frac{\op{k}_l}{\op{k}^2}-\av{z}^\pm_{j,ll}\frac{\op{k}_i}{\op{k}^2}.
\end{aligned}
    \end{gather}
\end{subequations}

From \Eq{eq:zintermsofw} and \Eq{eq:wintermsofz}, we have:
\begin{equation}\label{eq:zwrel}
    \ket{\fl{\vec{z}}}=\ii \op{k}^{-2}\bop{\mathrm{k}}_{\wedge}\ket{\fl{\vec{w}}},\qquad \ket{\fl{\vec{w}}}=\ii \bop{\mathrm{k}}_{\wedge}\ket{\fl{\vec{z}}},
\end{equation}
where 
\begin{equation}
    \bop{\mathrm{k}}_{\wedge}\doteq\begin{pmatrix}
        \bop{k}_{\wedge}&0\\0&\bop{k}_{\wedge}
    \end{pmatrix}, \qquad \bop{k}_{\wedge}\doteq\begin{pmatrix}
        0&-\op{k}_z&\op{k}_y\\
        \op{k}_z&0&-\op{k}_x\\
        -\op{k}_y&\op{k}_x&0
    \end{pmatrix}.
\end{equation}
Using \Eq{eq:zwrel} and \Eq{eq:fluc}, \Eq{eq:schrodinger} can then be put in a simpler form for the state vector \m{\ket{\vec{z}}}, with Hamiltonian \m{\bop{H}\doteq \ii\op{k}^{-2}\bop{\mathrm{k}}_{\wedge}\bop{H}'\bop{\mathrm{k}}_{\wedge}}. Specifically, \Eq{eq:schrodinger} becomes
\begin{equation}\label{eq:schrodinger2}
\ii\pd_t\ket{\fl{\vec{z}}}=\bop{H}\ket{\fl{\vec{z}}},
\end{equation}
where \m{\bop{H}} is a matrix operator given by 
\begin{equation}\label{eq:hamiltonian2}
   \bop{H}=\begin{pmatrix}  \bop{H}^{++}&\bop{H}^{+-}\\\bop{H}^{-+}&\bop{H}^{--}
            \end{pmatrix},
\end{equation}
and we have also introduced
\begin{subequations}
    \begin{gather}
\op{H}^{\pm\pm}_{ij}=\delta_{ij}\Big(\av{z}^\mp_l\op{k}_l-\ii\nu_+\op{k}^2\Big)+\ii\frac{\op{k}_i}{\op{k}^2}\av{z}_{l,j}^\mp\op{k}_l,\\
        \op{H}_{ij}^{\pm\mp}=-\delta_{ij}\ii\nu_-\op{k}^2-\ii\av{z}^\pm_{i,j}+\ii\frac{\op{k}_i}{\op{k}^2}\av{z}^\pm_{l,j}\op{k}_l.
    \end{gather}
\end{subequations}
Note that the Hamiltonian \m{\hat{\matr{H}}} is generally not Hermitian (even in the ideal-MHD limit) and is highly nontrivial. Also, because \m{\ket{\vec{z}}} is a multi-component vector, one can think of the quasiparticles as quantumlike particles with spin, and the `spin-up' and `spin-down' components are strongly coupled in the presence of inhomogeneous mean fields. 

In principle, \m{\hat{\matr{H}}} can be diagonalised for two-dimensional (2-D) dynamics when \m{|\av{\vec{b}}|\gg|\av{\vec{v}}|} (\App{app:2d}), such that the corresponding dynamics can be understood in terms of the resulting phase-space trajectories available to quasiparticles. A more detailed investigation of the properties of the Hamiltonian \Eq{eq:hamiltonian2} may be of interest for future work and may reveal ways to make greater use of the quasiparticle analogy.\footnote{For example, the topology of phase-space trajectories available to drift-wave quasiparticles has been used to explain the nonlinear saturation dynamics of zonal flows by \citet{ref:zhu19}.} However, such an exploration is beyond the scope of this paper. Also, for 3-D dynamics that is of interest in the context of the dynamo problem, a diagonalisation of \m{\hat{\matr{H}}} does not seem possible.

\subsubsection{Wigner--Moyal equation}
Right-multiplying \Eq{eq:schrodinger} by \m{\bra{\fl{\vec{z}}}} and subtracting the adjoint of the resulting equation yields the von-Neumann equation for the density operator of Alfv\'enic fluctuations, \m{\bop{W}\doteq\ket{\fl{\vec{z}}}\!\bra{\fl{\vec{z}}}}:
\begin{equation}\label{eq:neumann}
    \ii\pd_t\bop{W}=\bop{H}\bop{W}-\bop{W}\bop{H}^\dagger.
\end{equation}
Applying the Wigner--Weyl transform (\App{app:wignerweyl}) to \Eq{eq:neumann} yields the Wigner--Moyal equation (WME), which governs the dynamics of Alfv\'enic fluctuations in the phase space \m{(\vec{x},\vec{k})}:
    \begin{gather}\label{eq:unavwme}
        \ii\pd_t\matr{W}=\matr{H}\star\matr{W}-\matr{W}\star\matr{H}^\dagger,
    \end{gather}
where \m{\matr{H}} is the symbol of the Hamiltonian \m{\boper{H}}:
        \begin{gather}
            \matr{H}=\begin{pmatrix}
                \matr{H}^{++}&\matr{H}^{+-}\\
                \matr{H}^{-+}&\matr{H}^{--}
            \end{pmatrix},         
    \end{gather}
and the individual components are given by
\begin{subequations}
    \begin{gather}
H^{\pm\pm}_{ij}=\delta_{ij}\Big(\av{z}^\mp_l\star k_l-\ii\nu_+k^2\Big)+\ii\frac{k_i}{k^2}\star\av{z}_{l,j}^\mp\star k_l,\\
        H_{ij}^{\pm\mp}=-\delta_{ij}\ii\nu_-k^2-\ii\av{z}^\pm_{i,j}+\ii\frac{k_i}{k^2}\star\av{z}_{l,j}^\mp\star k_l,
    \end{gather}
\end{subequations}
the Moyal star product \m{\star} \Eq{eq:moyalstar} is defined in \App{app:wignerweyl}, and \m{\matr{W}} is the symbol of \m{\boper{W}}, also known as the Wigner matrix. The latter can be expressed as
\begin{equation}\label{eq:noavwigner}
     \matr{W}=\begin{pmatrix}
        \matr{W}^{++}&\matr{W}^{+-}\\
        \matr{W}^{-+}&\matr{W}^{--}
    \end{pmatrix},
\end{equation}
where
\begin{equation}
\begin{gathered}
        W^{\sigma_1\sigma_2}_{ij}=\int\dd \vec{s}\,\ee^{-\ii\vec{k}\cdot\vec{s}}\fl{z}^{\sigma_1}_i(\vec{x}+\vec{s}/2)\fl{z}^{\sigma_2}_j(\vec{x}-\vec{s}/2).
\end{gathered}
\end{equation}
Note that the Wigner matrix \m{\matr{W}} is Hermitian, \ie
\begin{equation}\label{eq:hermiticity}
    W_{ij}^{\sigma_1\sigma_2}(t,\vec{x},\vec{k})=W_{ji}^{\sigma_2\sigma_1*}(t,\vec{x},\vec{k}).
\end{equation} 
Additionally, since the Els\"asser fields are real, we also have that
\begin{equation}\label{eq:negk}
    W_{ij}^{\sigma_1\sigma_2}(t,\vec{x},\vec{k})=W_{ji}^{\sigma_2\sigma_1}(t,\vec{x},-\vec{k}).
\end{equation}
%

\subsubsection{Average Wigner matrix and the geometrical-optics expansion}

By averaging \Eq{eq:unavwme} with the same averaging operation used to define the mean fields, we obtain the averaged WME:
  \begin{gather}\label{eq:avwme}
        \ii\pd_t\av{\matr{W}}=\av{\matr{H}}\star\av{\matr{W}}-\av{\matr{W}}\star\av{\matr{H}}^\dagger.
    \end{gather}
Here, \m{\av{\matr{H}}=\matr{H}}, since \m{\matr{H}} is independent of the fluctuating fields, and the averaged Wigner matrix is the average of \m{\matr{W}},
\begin{equation}\label{eq:avwigner}
        \av{\matr{W}}\equiv\flav{\matr{W}}=\begin{pmatrix}
        \av{\matr{W}}^{++}&\av{\matr{W}}^{+-}\\
        \av{\matr{W}}^{-+}&\av{\matr{W}}^{--}
    \end{pmatrix},
    \end{equation}
    where
    \begin{equation}
        \av{W}^{\sigma_1\sigma_2}_{ij}=\int\dd \vec{s}\ee^{-\ii\vec{k}\cdot\vec{s}}\flav{\fl{z}^{\sigma_1}_i(\vec{x}+\vec{s}/2)\fl{z}^{\sigma_2}_j(\vec{x}-\vec{s}/2)}.
    \end{equation}
Notice that the average Wigner matrix \Eq{eq:avwigner} can be understood as the Fourier transform of the symmetrised two-point correlation tensor of the Els\"asser fields.
    
Below, the original Wigner matrix \Eq{eq:noavwigner} will not be needed, so we will call \Eq{eq:avwigner} `the' Wigner matrix and omit the bar in \m{\av{\matr{W}}} to simplify notation. The properties \Eq{eq:hermiticity} and \Eq{eq:negk} hold for this averaged matrix as well. Also, the trace of \m{\matr{W}}, to some extent\footnote{As discussed in \Sec{sec:basicnot}, the quasiparticle analogy for MHD in the absence of a strong guide field has its limitations due to the lack of a general diagonalisation for \m{\bop{H}}.}, can be interpreted as the phase-space density of turbulent quasiparticles.  In this sense, the WME can be considered as a generalisation of the Liouville equation. Its interpretation as a quantum extension of kinetic theory becomes clearer when we consider the series expansion of the Moyal star:
\begin{equation}\label{eq:wmeexpand}
    \begin{aligned}
        \ii\pd_t \matr{W} &=\matr{H}\ee^{\ii\op{\mc{L}}/2}\matr{W}-\matr{W}\ee^{\ii\op{\mc{L}}/2}\matr{H}\dagger\\
        &=\matr{H}(1+\ii\op{\mc{L}}/2+\hdots)\matr{W}-\matr{W}(1+\ii\op{\mc{L}}/2+\hdots)\matr{H}^\dagger,\\
    \end{aligned}
\end{equation}
where the Janus operator \m{\op{\mc{L}}} \Eq{eq:janusop} is defined in \App{app:wignerweyl}, and, basically, stands for the canonical Poisson bracket in the \m{(\vec{x},\vec{k})} space, $A\op{\mc{L}}B = \{A, B\}$. Note that \m{\op{\mc{L}}} effectively scales as the GO parameter \m{l/L}, so the higher-order terms (denoted `...' in \Eq{eq:wmeexpand}) can be omitted. Furthermore, this equation can be further simplified when W and H are scalar functions (as can be the case, for example, in drift-wave turbulence \citep{ref:zhu21}). In this case, \Eq{eq:wmeexpand} becomes the familiar `classical' wave kinetic equation when \m{l/L \to 0}:
\begin{equation}
    \pd_t W\approx\{H_{\mathrm{H}},W\}+2 H_{\mathrm{A}} W,
\end{equation}
where \m{H_{\mathrm{H}}} and \m{H_{\mathrm{A
}}} are the real (Hermitian) and the imaginary (anti-Hermitian) parts of the Hamiltonian.
However, keep in mind that, for MHD turbulence, \m{H} and \m{W} are non-commuting matrices. Furthermore, when \m{L} is determined by modulational instabilities, it is often the case that \m{l\sim L}, so the GO approximation of the WME is inapplicable.

\subsection{Turbulent source term for the mean field}
As with any quadratic functional of the fluctuating field \citep{my:ql}, the turbulent source term for the mean field,
\begin{equation}
    \vec{S}^\pm\doteq-\flav{\nabla\times[(\fl{\vec{z}}^\mp\cdot\nabla)\fl{\vec{z}}^\pm]},
\end{equation}
can be expressed through the Wigner matrix \m{\matr{W}}:
\begin{equation}
\begin{aligned}
        S_i^\pm=&-\flav{\epsilon_{ijk}\pd_{j}(\fl{z}^\mp_l\pd_l\fl{z}^\pm_k)}\\
        =&-\epsilon_{ijk}\flav{(\braket{\vec{x}|\ii\op{k}_j|\fl{z}^\mp_l}\!\braket{\vec{x}|\ii\op{k}_l|\fl{z}^\pm_k}-\braket{\vec{x}|\fl{z}^\mp_l}\!\braket{\vec{x}|\op{k}_j\op{k}_l|\fl{z}^\pm_k})}\\                =&-\epsilon_{ijk}\flav{(\braket{\vec{x}|\op{k}_j|\fl{z}^\mp_l}\!\braket{\fl{z}^\pm_k|\op{k}_l|\vec{x}}-\braket{\vec{x}|\fl{z}^\mp_l}\!\braket{\fl{z}^\pm_k|\op{k}_j\op{k}_l|\vec{x}})}\\
        =&\,\epsilon_{ijk}\int\frac{\dd\vec{k}}{(2\pi)^3}\Big(k_lk_j\star W_{kl}^{\pm\mp}-k_l\star W_{kl}^{\pm\mp}\star k_j\Big),
\end{aligned}
\end{equation}
where \m{\epsilon_{ijk}} is the Levi--Civita symbol and the transition from the third line to the fourth line uses \Eq{eq:useful}.

Note that the source term involves only the off-diagonal blocks of the Wigner matrix, \m{\matr{W}^{\pm\mp}}, \ie mean fields are generated only by correlations between \m{\fl{\vec{z}}^+} and \m{\fl{\vec{z}}^-}. This can also be expected from the original Els\"asser equations \Eq{eq:elsasser}, where the nonlinear term vanishes if either \m{\vec{z}^+} or \m{\vec{z}^-} is zero.

\subsection{Summary of the main equations}
In summary, our mean-field wave-kinetics model is as follows:
\begin{subequations}\label{eq:summary}
\begin{gather}
        \pd_t \av{\vec{w}}^\pm = 
-\bop{k} \times
\Big\{
\Big[\Big(\frac{\bop{k}}{\op{k}^2} \times\av{\vec{w}}^\mp\Big)\cdot \bop{k}\Big]
\Big(\frac{\bop{k}}{\op{k}^2}\times\av{\vec{w}}^\pm\Big)
\Big\}
-\op{k}^2(\nu_+ \av{\vec{w}}^\pm + \nu_- \av{\vec{w}}^\mp)+\vec{S}^\pm
,\\
    \ii\pd_t\matr{W}=\matr{H}\star\matr{W}-\matr{W}\star\matr{H}^\dagger,\label{eq:WME}
    \end{gather}
\end{subequations}
where
\begin{equation}
            S_i^\pm=\epsilon_{ijk}\int\frac{\dd\vec{k}}{(2\pi)^3}\Big(k_lk_j\star W_{kl}^{\pm\mp}-k_l\star W_{kl}^{\pm\mp}\star k_j\Big),
\end{equation}
and the matrix
\begin{equation}
         \matr{H}=\begin{pmatrix}
                \matr{H}^{++}&\matr{H}^{+-}\\
                \matr{H}^{-+}&\matr{H}^{--}
            \end{pmatrix}
\end{equation}
consists of
\begin{subequations}
        \begin{gather}
             H^{\pm\pm}_{ij}=\delta_{ij}\Big(\av{z}^\mp_l\star k_l-\ii\nu_+k^2\Big)+\ii\frac{k_i}{k^2}\star\av{z}_{l,j}^\mp\star k_l,\\
        H_{ij}^{\pm\mp}=-\delta_{ij}\ii\nu_-k^2-\ii\av{z}^\pm_{i,j}+\ii\frac{k_i}{k^2}\star\av{z}_{l,j}^\pm\star k_l.
        \end{gather}
\end{subequations}
Note that \Eqs{eq:summary} are equivalent to the original MHD system within the QLA and do not assume scale separation between the fluctuations and mean fields, which is typically done in the literature \citep{ref:brandenburg18,ref:hughes18}. As we show in \App{app:conservationlaws}, equations \Eqs{eq:summary} conserve the total energy and cross-helicity, while allowing for the transfer of these invariants between the mean and fluctuating fields.

To aid comparisons with existing theories, we will also often work with the following ad hoc modification of \Eq{eq:WME}:
\begin{equation}\label{eq:modWME}
        \ii\pd_t\matr{W}=\matr{H}\star\matr{W}-\matr{W}\star\matr{H}^\dagger-\ii\tau_c^{-1}\matr{W}+\matr{T},
\end{equation}
where \m{\tau_c} is the correlation time that determines the damping of fluctuations through wave--wave collisions. The forcing term \m{\matr{T}}, which is yet to specified, is added to allow for turbulent equilibria at nonzero dissipation (\Sec{sec:linmod}). A similar term appears in the minimal tau approximation, or MTA \citep{ref:blackman02,ref:brandenburg05}.

\section{Linear modulational dynamics within MFWK\label{sec:linmod}}
Let us now consider small mean-field perturbations to an otherwise homogeneous turbulent background. This corresponds to 
\begin{gather}
    \matr{W}=(2\pi)^3[\matr{F}(\vec{k})+\matr{f}(t,\vec{x},\vec{k})],
\end{gather}
where \m{\matr{F}(\vec{k})} is the Wigner matrix of the homogeneous turbulent equilibrium, and \m{\matr{f}(t,\vec{x},\vec{k})} is the first-order response of the turbulence to the mean fields. (The factor $(2\pi)^3$ is introduced to shorten notation in some formulas below.) Similarly, the Hamiltonian will be of the form 
\begin{gather}
\matr{H}=\matr{H}_0(\vec{k})+\matr{h}(t,\vec{x},\vec{k}),
\end{gather} 
where \m{\matr{H}_0} captures viscous dissipation and resistivity, and \m{\matr{h}} is the perturbed Hamiltonian due to the mean fields.
\subsection{Properties of statistically homogeneous turbulent equilibria\label{sec:equilibria}}

\subsubsection{Equilibrium condition}
In the absence of zeroth-order mean fields, equations \Eq{eq:summary} give the following equations for the equilibrium:
\begin{subequations}\label{eq:0th}
\begin{gather}
       0 =\vec{S}_0^\pm,\\
    0=(2\pi)^3(\matr{H}_0\matr{F}-\matr{F}\matr{H}_0^\dagger-\ii\tau_{c}^{-1}\matr{F})+\matr{T},\\
    S_{0i}^\pm=\epsilon_{ijk}\int\dd\vec{k}(k_lk_j\star F_{kl}^{\pm\mp}-k_l\star F_{kl}^{\pm\mp}\star k_j),\\
    \begin{gathered}
            \matr{H}_{0}^{\pm\pm}=-\ii\nu_+k^2\matr{1}_3,\qquad
    \matr{H}_{0}^{\pm\mp}=-\ii\nu_-k^2\matr{1}_3.\\
    \end{gathered}
\end{gather}
\end{subequations}
where \m{\matr{1}_3} is the \m{3\times 3} identity matrix.
Thus, for a given \m{\matr{F}}, the matrix \m{\matr{T}} must satisfy
\begin{equation}\label{eq:tcond}
    \matr{T}^{\sigma_1\sigma_2}=\ii(2\pi)^3\Big[(2\nu_+ k^2+\tau_c^{-1})\matr{F}^{\sigma_1\sigma_2}+\nu_-k^2(\matr{F}^{\sigma_1\overbar{\sigma_2}}+\matr{F}^{\overbar{\sigma_1}\sigma_2})\Big],
\end{equation}
where \m{\sigma_{1/2}=+/-} and \m{\overbar{\sigma}\doteq -\sigma}.

Note that in the absence of dissipation (\m{\tau_c^{-1}=0} and \m{\nu_\pm=0}), \Eq{eq:tcond} gives \m{\matr{T}=0}; \ie any such homogeneous turbulent background is quasilinearly self-consistent without external driving. Also note that the consistency of the mean-field equation, \m{\vec{S}_0=0}, follows directly from the background homogeneity, \m{\matr{F}(\vec{x},\vec{k})=\matr{F}(\vec{k})}:
\begin{equation}
    \begin{aligned}
        S_{0i}^\pm=\,&\epsilon_{ijk}\int\dd\vec{ k}\,(k_lk_j\star F_{kl}^{\pm\mp}-k_l\star F_{kl}^{\pm\mp}\star k_j)\\
        =\,&\epsilon_{ijk}\int\dd \vec{k}\,(k_lk_j-k_l k_j) F_{kl}^{\pm\mp}\\
        =\,&0.
    \end{aligned}
\end{equation}
%

\subsubsection{Wigner matrix of isotropic MHD turbulence}
For a homogeneous turbulent background, we have
\begin{equation}
\begin{aligned}
         F_{ij}^{\sigma_1\sigma_2}(\vec{k})&=\frac{1}{(2\pi)^3}\int\dd\vec{ s}\,\ee^{-\ii\vec{k}\cdot\vec{s}}\Big\langle\fl{z}_i^{\sigma_1}\Big(\vec{x}+\frac{\vec{s}}{2}\Big)\fl{z}_j^{\sigma_2}\Big(\vec{x}-\frac{\vec{s}}{2}\Big)\Big\rangle\\
         &=R_{ij}^{\sigma_1\sigma_2}(-\vec{k}),
\end{aligned}
\end{equation}
where \m{\matr{R}^{\sigma_1\sigma_2}(\vec{k})} is the Fourier transform of the two-point correlation tensor:
\begin{equation}
    R_{ij}^{\sigma_1\sigma_2}(\vec{r})=\flav{\fl{z}^{\sigma_1}_i(\vec{x}-\vec{r}/2)\fl{z}^{\sigma_2}_j(\vec{x}+\vec{r}/2)}.
\end{equation}

If we further assume that the turbulence is isotropic \citep{ref:oughton97}, then
\begin{equation}
    R_{ij}^{\sigma_1\sigma_2}(\vec{k})=\Big(\delta_{ij}-\frac{k_i k_j}{k^2}\Big)\frac{E^{\sigma_1\sigma_2}(k)}{4\pi k^2}+\ii \epsilon_{ijk}\frac{k_k}{k^2}\frac{H^{\sigma_1\sigma_2}(k)}{8\pi k^2},
\end{equation}
where \m{\matr{R}^{\sigma_1\sigma_2}=\matr{R}^{\sigma_2\sigma_1}}, \m{k\doteq|\vec{k}|}, and the \m{H^{\sigma_1\sigma_2}} are zero for non-helical turbulence. Note that \m{E^{\sigma_1\sigma_2}(k)} and \m{H^{\sigma_1\sigma_2}(k)} are the spectra of energy-like and helicity-like quantities respectively, in the following sense:
\begin{equation}
    \begin{gathered}
        \flav{\vec{z}^\sigma_1\cdot\vec{z}^\sigma_2}=2\int\dd k \,E^{\sigma_1\sigma_2}(k),\qquad \flav{\vec{z}^\sigma_1\cdot(\nabla\times\vec{z}^\sigma_2)}=\int\dd k \,H^{\sigma_1\sigma_2}(k).
    \end{gathered}
\end{equation}
In other words, isotropic MHD turbulence corresponds to
\begin{equation}\label{eq:isowigner}
\begin{gathered}
        \matr{F}=\begin{pmatrix}
        \matr{F}^+&\matr{F}^C\\
        \matr{F}^C&\matr{F}^-
    \end{pmatrix},\\
    F^\chi_{ij}(\vec{k})=\Big(\delta_{ij}-\frac{k_i k_j}{k^2}\Big)\frac{E^\chi(k)}{4\pi k^2}-\ii \epsilon_{ijk}\frac{k_k}{k^2}\frac{H^\chi(k)}{8\pi k^2}
\end{gathered}
\end{equation}
with \m{\chi=+,\,-,\,C}. It will also be convenient to work with the symmetric and antisymmetric combinations \m{\matr{F}^S\doteq(\matr{F}^+ +\matr{F}^-)/2} and \m{\matr{F}^A\doteq (\matr{F}^+ -\matr{F}^-)/2}, and we also extend this notation to \m{E} and \m{H}. Then,
\begin{equation}\label{eq:spectra}
    \begin{gathered}
      \int_0^\infty\dd k \,E^{S}(k)=\frac{1}{2}\flav{\vec{\fl{v}}\cdot\vec{\fl{v}}+\vec{\fl{b}}\cdot\vec{\fl{b}}},\\
    \int_0^\infty\dd k \,E^{C}(k)=\frac{1}{2}\flav{\vec{\fl{v}}\cdot\vec{\fl{v}}-\vec{\fl{b}}\cdot\vec{\fl{b}}},\\
    \int_0^\infty\dd k \,E^{A}(k)=\flav{\vec{\fl{v}}\cdot\vec{\fl{b}}},\\
        \int_0^\infty\dd k \,H^{S}(k)=\flav{\vec{\fl{v}}\cdot\fl{\vec{w}}+\vec{\fl{b}}\cdot\fl{\vec{j}}},\\
    \int_0^\infty\dd k \,H^{C}(k)=\flav{\vec{\fl{v}}\cdot\fl{\vec{w}}-\vec{\fl{b}}\cdot\fl{\vec{j}}},\\
    \int_0^\infty\dd k \,H^{A}(k)=2\flav{\vec{\fl{v}}\cdot\fl{\vec{j}}}=2\flav{\fl{\vec{w}}\cdot\fl{\vec{b}}},
    \end{gathered}
\end{equation}
where \m{\fl{\vec{w}}\doteq\nabla\times\fl{\vec{v}}} and \m{\fl{\vec{j}}\doteq\nabla\times\fl{\vec{b}}}. 

\subsection{Linear modulational dynamics}
\subsubsection{Linearised equations}
The perturbed quantities are governed by the following linearised MFWK equations:
\begin{subequations}\label{eq:linearised}
\begin{gather}
        \pd_t \av{\vec{w}}^\pm =
-\op{k}^2(\nu_+ \av{\vec{w}}^\pm + \nu_- \av{\vec{w}}^\mp)+\vec{s}^\pm
,\\
    \ii\pd_t\matr{f}=\matr{H_0}\star\matr{f}-\matr{f}\star\matr{H_0}^\dagger+\matr{h}\star\matr{F}-\matr{F}\star\matr{h}^\dagger-\ii\tau_c^{-1}\matr{f},\\
    s_i^\pm=\epsilon_{ijk}\int\dd\vec{k}\,(k_lk_j\star f_{kl}^{\pm\mp}-k_l\star f_{kl}^{\pm\mp}\star k_j),\\
        \begin{gathered}
h^{\pm\pm}_{ij}=\delta_{ij}\Big(\av{z}^\mp_l\star k_l\Big)+\ii\,\frac{k_i}{k^2}\star\av{z}_{l,j}^\mp\star k_l,\\
        h_{ij}^{\pm\mp}=-\ii\av{z}^\pm_{i,j}+\ii\,\frac{k_i}{k^2}\star\av{z}_{l,j}^\pm\star k_l.
        \end{gathered}
\end{gather}
\end{subequations}
Note that this model is fundamentally different from the commonly used kinematic approximation \citep{book:krause80}. The kinematic approximation assumes that all magnetic fields (that is, both turbulent and mean components) are sufficiently weak such that the flow (again, both mean and turbulent parts) can be taken as prescribed. The momentum equation is then never invoked, and the resulting mean-field theory is built on the induction equation alone. It has already been pointed out in the literature \citep{ref:courvoisier10,ref:courvoisier10b} that such an approach, which artificially privileges the magnetic field over the flow, is fundamentally inconsistent in the presence of substantial magnetic field fluctuations (which are to be expected in MHD turbulence), but the general theory for this case has been lacking. Our approach fixes this problem in that it is formulated in terms of the Els\"asser fields and thus treats velocity and magnetic fields on the same footing.

\subsubsection{Eikonal perturbations}
Let us now look for the linear eigenmodes of \Eq{eq:linearised}. For that, let us consider perturbed quantities of the following form:
\begin{subequations}\label{eq:eikonalform}
\begin{gather}
               \av{\vec{z}}^\pm=\re \left(\vec{\pz}^\pm\ee^{\ii\Theta}\right),\label{eq:eikonalz}\\
               \av{\vec{w}}^\pm=\re \left(\vec{\pw}^\pm\ee^{\ii\Theta}\right),\\
        \matr{f}=\left(\matr{\pf}(\vec{k})\ee^{\ii\Theta}\right)_{\mathrm{H}},\\ \matr{h}=\frac{1}{2}(\matr{\ph}_{(+)}(\vec{k})\ee^{i\Theta}+\matr{\ph}_{(-)}^{\dagger}(\vec{k})\ee^{-i\Theta^*}),
\end{gather}
\end{subequations}
with
\begin{equation}\label{eq:modphase}
    \Theta=-\Omega t+\vec{K}\cdot\vec{x},
\end{equation}
where \m{\Omega} and \m{\vec{K}} are the modulational frequency and wavevector respectively, and \m{\vec{\pw}^\pm=\ii\vec{K}\times \vec{\pz}^\pm}. (The index H denotes the Hermitian part.) We will use the convention that the modulational wavevector \m{\vec{K}} is real, while the modulational frequency \m{\Omega} may be complex.
Note that the mean-field polarizations, \m{\vec{\pz}^\pm} and \m{\vec{\pw}^\pm}, are constants, while \m{\matr{\ph}_{(\pm)}} and \m{\matr{\pf}} are functions of \m{\vec{k}}. We also use the convention that the argument of a function will only be explicitly written when it is first defined, not obvious from the context, or judged to provide helpful information.

\subsubsection{Equations for the polarizations}
Using \Eq{eq:starshift}, we can write the following equations for the polarizations \m{\vec{\pz}^\pm} and \m{\matr{\pf}}:

\begin{gather}\label{eq:polmf}
\Omega \vec{\pz}^\pm=-\ii K^2(\nu_+ \vec{\pz}^\pm+\nu_-\vec{\pz}^\mp)-\frac{\vec{K}}{K^2}\times\vec{s}^\pm,\\
\begin{aligned}\label{eq:polWME}
    \Omega'\matr{\pf}-\matr{H}_0(\vec{k}_+)\matr{\pf}+\matr{\pf}\matr{H}_0^{\dagger}(\vec{k}_-)=\matr{\ph}_{(+)}\matr{F}(\vec{k}_-)-\matr{F}(\vec{k}_+)\matr{\ph}_{(-)},
\end{aligned}
\end{gather}
where
\begin{gather}
    -\Big(\frac{\vec{K}}{K^2}\times\vec{s}^\pm\Big)_i=\int\dd\vec{k}\Big(K_n\pf_{in}^{\pm\mp}-\frac{K_iK_mK_n}{K^2}\pf_{mn}^{\pm\mp}\Big),
\end{gather}
and
\begin{gather}
\matr{\ph}_{(\pm)}=\begin{pmatrix}
    \matr{\ph}_{(\pm)}^{++}&\matr{\ph}_{(\pm)}^{+-}\\
    \matr{\ph}_{(\pm)}^{-+}&\matr{\ph}_{(\pm)}^{--}
\end{pmatrix},
\end{gather}
where
\begin{subequations}
    \begin{gather}
         \ph^{\pm\pm}_{(+)ij}=(\vec{\pz}^\mp\cdot\vec{k})\Big(\delta_{ij}-\frac{k_{+i}K_j}{k_+^2}\Big), \\
         \ph^{\pm\mp}_{(+)ij}=\,\pz_i^\mp K_j-(\vec{\pz}^\mp\cdot\vec{k})\frac{k_{+i}K_j}{k_+^2},\\
    \,\ph^{\pm\pm}_{(-)ij}=(\vec{\pz}^\mp\cdot\vec{k})\Big(\delta_{ij}+\frac{K_ik_{-j}}{k_-^2}\Big), \\\ph^{\pm\mp}_{(-)ij}= -\pz_j^\pm K_i+(\vec{\pz}^\pm\cdot\vec{k})\frac{K_i k_{-j}}{k_-^2}.
    \end{gather}
\end{subequations}
Also, \m{\Omega'\doteq\Omega+\ii\tau_c^{-1}}, \m{\vec{k}_\pm\doteq\vec{k}\pm\vec{K}/2}, and \m{k_{\pm i}} refers to the \m{i}th element of \m{\vec{k}_\pm}. We study applications of these equations in \Sec{sec:nonlocalemf} and \Sec{sec:modmodes}.

\section{Nonlocal turbulent EMF\label{sec:nonlocalemf}}
Much of mean-field dynamo theory is concerned with the calculation of the turbulent EMF \citep{ref:squire15_a,ref:yokoi18,ref:rogachevskii03,ref:radler10,ref:radler07}. One of the primary limitations of traditional mean-field theories is that they assume scale separation, i.e. the existing analytical closures are local in the sense that the EMF at a given point in space-time is expressed purely in terms of the magnetic field and its low-order derivatives at that same point.

Scale separation is often poorly justified, and nonlocality may play an important role in astrophysical systems of interest \citep{ref:kapyla06, ref:brandenburg18}. In such cases, the local formulation of the EMF must be replaced with the nonlocal-response kernel:
\begin{equation}
    \vec{\emf}(t,\vec{x})=\int\dd t'\int\dd \vec{x}' \,\matr{G}(t,\vec{x};t',\vec{x}')\,\av{\vec{b}}(t',\vec{x}').
\end{equation}
When considering small mean-field based departures from an otherwise homogeneous turbulent equilibrium, such that \m{\matr{G}(t,\vec{x};t',\vec{x}')=\matr{G}(t-t',\vec{x}-\vec{x}')}, this can be expressed as a simple product of the Fourier images of \m{\matr{G}} and \m{\av{\vec{b}}}:
\begin{equation}\label{eq:fourieremf}
    \vec{\emf}(\Omega,\vec{K})=\matr{G}(\Omega,\vec{K})\av{\vec{b}}(\Omega,\vec{K}).
\end{equation}
Note that we only explicitly included the dependence of the EMF on the mean magnetic field in the discussion above, as this is the case that has been considered in the literature. As we shall see shortly, neglecting contributions from the mean velocity field is only valid for purely hydrodynamic turbulent backgrounds.

Determining \m{\matr{G}(\Omega,\vec{K})} is the key to incorporating nonlocality in mean-field models and is the subject of ongoing research \citep{ref:pipin23,ref:bendre22,ref:rheinhardt14,ref:rheinhardt12,ref:hubbard09,ref:brandenburg08,ref:gressel20}. Current efforts to probe nonlocality typically use the `test-field method' \citep{ref:schrinner05,ref:schrinner07}, in which various test mean fields are imposed on some turbulent background, and the induction equation for the fluctuating magnetic field is solved, allowing for transport coefficients to be inferred numerically. Such efforts have deduced the following approximate form for the nonlocal response kernel:
\begin{equation}\label{eq:numkernel}
    G_{ij}(\Omega,\vec{K})\approx\frac{\alpha+\ii\epsilon_{ijk}K_k \beta}{1-\ii\tau\Omega+l^2 K^2 },
\end{equation}
where
\begin{equation}\label{eq:localab}
    \alpha=-\frac{\tau_c}{3}\flav{\fl{\vec{v}}\cdot\nabla\times\fl{\vec{v}}},\qquad\beta=\frac{\tau_c}{3}\flav{\fl{\vec{v}}\cdot\fl{\vec{v}}}
\end{equation}
are the local turbulent transport coefficients corresponding to the \m{\alpha}-effect and turbulent diffusivity, respectively. The coefficients \m{\tau} and \m{l} are fitting parameters that represent the degree of nonlocality, and while \m{\tau} is consistently found to be on the order of the eddy-turnover time, there is less, if any, agreement in the literature on how to understand and model $l$ \citep{ref:rheinhardt12, ref:brandenburg18}.
Advancing our understanding of this nonlocality requires an analytical prediction for the nonlocal response kernel, which is not possible with the traditional local mean-field formalism.

MFWK enables an analytic calculation of the nonlocal-response kernel from first principles (within the QLA), and in the weakly inhomogeneous limit. This result serves as a first step in providing physical motivation for the form of the nonlocal response kernel. By comparisons with test field methods \citep{ref:schrinner05,ref:rheinhardt12}, it can help isolate which aspects of nonlocality can be fully described within mean-field effects, and which are fundamentally nonlinear. In the remainder of this section, we outline the calculation and highlight main results.

\subsection{Derivation\label{sec:nonlocalemfderiv}}
The generic nonlocal linear response of the turbulent EMF \m{\vec{\emf}} to the mean fields \m{\av{\vec{b}}} and \m{\av{\vec{v}}} can be written as follows:
\begin{equation}
    \vec{\emf}(t,\vec{x})=\int\dd t'\int\dd\vec{x}'\Big(\matr{G}^{(\vec{b})}(t,x;t',x')\,\av{\vec{b}}(t',x')+\matr{G}^{(\vec{v})}(t,x;t',x')\,\av{\vec{v}}(t',x')\Big).
\end{equation}
Note that we are allowing for a possible dependence on the mean flow, which is often neglected. (Since the mean flow is generally inhomogeneous, for example, as in \Eq{eq:eikonalz}, it cannot be removed by a Galilean transformation.) For the remainder of our calculation of the turbulent EMF, we will also work with the traditional velocity and magnetic fields, as opposed to the Els\"asser fields, in order to aid comparisons with existing theories.

We now assume weakly inhomogeneous turbulence, such that \m{\av{\vec{z}}^\pm} and \m{\av{\vec{w}}^\pm} are perturbations to an otherwise homogeneous turbulent background. In this case, we can take \m{\matr{G}^{(\vec{b},\vec{v})}(t,x;t',x')=\matr{G}^{(\vec{b},\vec{v})}(t-t',x-x')} such that in Fourier space we have:
\begin{equation} \label{eq:greensfuncform}
    \vec{\emf}(\Omega,\vec{K})=\matr{G}^{(\vec{b})}(\Omega,\vec{K})\av{\vec{b}}(\Omega,\vec{K})+\matr{G}^{(\vec{v})}(\Omega,\vec{K})\av{\vec{v}}(\Omega,\vec{K}).
\end{equation}
(For the remainder of our calculation of the turbulent EMF, we will also work with the traditional velocity and magnetic fields, as opposed to the Els\"asser fields, in order to aid comparisons with existing theories.) An expression relating the Fourier coefficients of \m{\vec{\emf}}, \m{\av{\vec{b}}} and \m{\av{\vec{v}}} at frequency \m{\Omega} and wavenumber \m{\vec{K}} therefore yields the nonlocal response kernel \m{\matr{G}(\Omega,\vec{K})}. 

To do this, we first note that the EMF can be directly written in terms of our Wigner matrix as follows:
\begin{equation}\label{eq:emfwigner}
    \begin{aligned}
        \emf_i&=-\frac{1}{2}\flav{\fl{\vec{z}}^+\times\fl{\vec{z}}^-}_i\\
        &=-\frac{1}{2}\int\frac{\dd\vec{k}}{(2\pi)^3}\, \epsilon_{ijk}W^{+-}_{jk}.
\end{aligned}
\end{equation}
Curiously, \Eq{eq:emfwigner} can also be expressed as
\begin{equation}
    \begin{gathered}
        \vec{\emf}=\bigstar\int\frac{\dd\vec{k}}{(2\pi)^3}\,\matr{W}^{-+},
\end{gathered}
\end{equation}
where \m{\bigstar} is the Hodge star, \m{(\bigstar \vec{A})_i\doteq\epsilon_{ijk}A_{jk}/2}.
The desired expression can therefore be obtained with the modulational mode analysis described in \Sec{sec:linmod}, by solving the WME to obtain the Wigner matrix in terms of the mean fields (which enter the WME through \m{\matr{H}}). 

 We can immediately see that there may, in principle, be some contribution to the turbulent EMF from \m{\matr{F}}, which has nothing to do with the mean fields:
\begin{equation}
    \begin{aligned}
        \emf_{0i}=&-\frac{\epsilon_{ijk}}{2}\int \dd\vec{k}\,F_{jk}^{+-}\\
        =&-\frac{\epsilon_{ijk}}{4}\int\dd\vec{ k}\,(F_{jk}^{+-}+F_{kj}^{-+})\\
        =&-\frac{\epsilon_{ijk}}{4}\int\dd\vec{ k}\,(F_{jk}^{+-}-F_{jk}^{-+}).\\
    \end{aligned}
\end{equation}
It can be easily shown that such a background EMF \m{\vec{\emf}_0} vanishes for a broad class of turbulent backgrounds.
For example, isotropy is a sufficient but not necessary condition for \m{\matr{F}^{+-}=\matr{F}^{-+}}. We henceforth understand \m{\vec{\emf}} to refer to the mean-field contribution alone and will not discuss \m{\vec{\emf}_0} further.

The mean-field contribution to the turbulent EMF is captured by the perturbed Wigner matrix:
\begin{equation}
         \emf_i=-\epsilon_{ijk}\int\dd\vec{k} f_{jk}^{+-}.
\end{equation}
As in \Sec{sec:linmod}, we now take our perturbed quantities to be of the following form:
\begin{equation}\label{eq:eikonalemf}
\begin{gathered}
 \av{\vec{v}}^\pm=\re \left(\vec{\pv}^\pm\ee^{\ii\Theta}\right),\qquad\av{\vec{b}}^\pm=\re \left(\vec{\pb}^\pm\ee^{\ii\Theta}\right),\qquad\vec{\emf}=\re\left(\vec{\pemf}\ee^{\ii\Theta}\right),\\
\matr{f}=\left(\matr{\pf}(\vec{k})\ee^{\ii\Theta}\right)_{H},\qquad \matr{h}=\frac{1}{2}(\matr{\ph}_{+}(\vec{k})\ee^{i\Theta}+\matr{\ph}_-^{\dagger}(\vec{k})\ee^{-i\Theta^*}),\\
\end{gathered}
\end{equation}
with \m{\Theta=-\Omega t+\vec{K}\cdot\vec{x}}, where \m{\Omega} and \m{\vec{K}} are the modulational frequency and wavevector, respectively. In terms of the notation of \Eq{eq:greensfuncform}, one has 
\begin{equation}
    \vec{\emf}(\Omega,\vec{K})=\vec{\pemf}, \qquad\av{\vec{b}}(\Omega,\vec{K})=\vec{\pb}, \qquad\av{\vec{v}}(\Omega,\vec{K})=\vec{\pv}.
\end{equation}
Therefore, our relevant equations are
\begin{gather}
        \pemf_i=-\frac{1}{2}\int\dd\vec{k} \,\epsilon_{ijk}\pf_{jk}^{+-},\label{eq:polEMF}
\end{gather}
along with \Eq{eq:polWME}, \m{\vec{\pv}=(\vec{\pz}^++\vec{\pz}^-)/2}, and \m{\vec{\pb}=(\vec{\pz}^+-\vec{\pz}^-)/2}. 

Solving \Eq{eq:polWME} for the off-diagonal blocks of the perturbed Wigner matrix yields
\begin{equation}\label{eq:offdiagf}
    \begin{aligned}
        \matr{\pf}^{\pm\mp}&=\\
        &\frac{C_1(\vec{k})}{C(\vec{k})}\Big[\matr{\ph}^{\pm+}_{(+)}\matr{F}^{+\mp}(\vec{k}_-)+\matr{\ph}^{\pm-}_{(+)}\matr{F}^{-\mp}(\vec{k}_-)-\matr{F}^{\pm+}(\vec{k}_+)\matr{\ph}^{+\mp}_{(-)}-\matr{F}^{\pm-}(\vec{k}_+)\matr{\ph}^{-\mp}_{(-)}\Big]\\
        +&\frac{C_2(\vec{k})}{C(\vec{k})}\Big[\matr{\ph}^{\mp+}_{(+)}\matr{F}^{+\pm}(\vec{k}_-)+\matr{\ph}^{\mp-}_{(+)}\matr{F}^{-\pm}(\vec{k}_-)-\matr{F}^{\mp+}(\vec{k}_+)\matr{\ph}^{+\pm}_{(-)}-\matr{F}^{\mp-}(\vec{k}_+)\matr{\ph}^{-\pm}_{(-)}\Big]\\
        +&\frac{C_3(\vec{k})}{C(\vec{k})}\Big[\matr{\ph}^{\mp+}_{(+)}\matr{F}^{+\mp}(\vec{k}_-)+\matr{\ph}^{\mp-}_{(+)}\matr{F}^{-\mp}(\vec{k}_-)-\matr{F}^{\mp+}(\vec{k}_+)\matr{\ph}^{+\mp}_{(-)}-\matr{F}^{\mp-}(\vec{k}_+)\matr{\ph}^{-\mp}_{(-)}\Big]\\
        +&\frac{C_4(\vec{k})}{C(\vec{k})}\Big[\matr{\ph}^{\pm+}_{(+)}\matr{F}^{+\pm}(\vec{k}_-)+\matr{\ph}^{\pm-}_{(+)}\matr{F}^{-\pm}(\vec{k}_-)-\matr{F}^{\pm+}(\vec{k}_+)\matr{\ph}^{+\pm}_{(-)}-\matr{F}^{\pm-}(\vec{k}_+)\matr{\ph}^{-\pm}_{(-)}\Big],
    \end{aligned}
\end{equation}
where
\begin{equation}\label{eq:thecs}
    \begin{gathered}
        C(\vec{k})=[\mc{W}^2(\Omega,\vec{K},\vec{k})+\nu_-^2( k_+^4+k_-^4)]^2-4\nu_-^4k_+^4k_-^4,\\
        C_1(\vec{k})=\mc{W}(\Omega,\vec{K},\vec{k})\Big[\mc{W}^2(\Omega,\vec{K},\vec{k})+\nu_-^2(k_+^4+k_-^4)\Big],\\
        C_2(\vec{k})=-2\nu_-^2k_+^2k_-^2\mc{W}(\Omega,\vec{K},\vec{k}),\\
        C_3(\vec{k})=-\ii\nu_-k_+^2\Big[\mc{W}^2(\Omega,\vec{K},\vec{k})+\nu_-^2( k_+^4+k_-^4)\Big]+2\ii\nu_-^3k_-^4k_+^2,\\
        C_4(\vec{k})=-\ii\nu_-k_-^2 \Big[\mc{W}^2(\Omega,\vec{K},\vec{k})+\nu_-^2( k_+^4+k_-^4)\Big]+2\ii\nu_-^3k_+^4k_-^2,\\
        \mc{W}(\Omega,\vec{K},\vec{k})\doteq\Omega'+\ii\nu_+\Big(k_+^2+k_-^2\Big),\\
    \end{gathered}
\end{equation}
and, as before, \m{\vec{k}_\pm=\vec{k}\pm\vec{K}/2}. Noting that
\begin{equation}\label{eq:theds}
    \begin{gathered}
         D_1(\vec{k})\doteq\frac{C_1(\vec{k}_+)}{C(\vec{k}_+)}=\frac{C_1(-\vec{k}_+)}{C(-\vec{k}_+)},\qquad
        D_2(\vec{k})\doteq\frac{C_2(\vec{k}_+)}{C(\vec{k}_+)}=\frac{C_2(-\vec{k}_+)}{C(-\vec{k}_+)},\\
         D_3(\vec{k})\doteq\frac{C_3(\vec{k}_+)}{C(\vec{k}_+)}=\frac{C_4(-\vec{k}_+)}{C(-\vec{k}_+)},\qquad
         D_4(\vec{k})\doteq\frac{C_4(\vec{k}_+)}{C(\vec{k}_+)}=\frac{C_3(-\vec{k}_+)}{C(-\vec{k}_+)},\\
    \end{gathered}
\end{equation}
let us define
\begin{equation}
    \matr{\ph}(\vec{k})\doteq\matr{\ph}_{(+)}(\vec{k}_+)=-\matr{\ph}_{(-)}^\intercal(-\vec{k}_+),
\end{equation}
where $^\intercal$ denotes transposition and
\begin{equation}\label{eq:polhs}
\begin{gathered}
      \ph_{ij}^{\pm\pm}(\vec{k})=(\vec{\pz}^\mp\cdot\vec{k})\Big[\delta_{ij}-\frac{(k_i+K_i)K_j}{(\vec{k}+\vec{K})^2}\Big],\\
    \ph_{ij}^{\pm\mp}(\vec{k})=\pz_i^\mp K_j-(\vec{\pz}^\mp\cdot\vec{k})\frac{(k_i+K_i)K_j}{(\vec{k}+\vec{K})^2}.
\end{gathered}
\end{equation}
Substituting \Eq{eq:offdiagf} into \Eq{eq:polEMF} yields:
\begin{equation}\label{eq:polEMFfinal}
\begin{aligned}
        \pemf_i=-\frac{1}{2}\int\vec{\dd}\vec{k}\,\epsilon_{ijk}\Big\{&-\Big[\Delta_1(\vec{k})\ph^{-+}_{jl}(\vec{k})+\Delta_2(\vec{k})\ph^{++}_{jl}(\vec{k})\Big]F^{++}_{lk}(\vec{k})\\
        &+\Big[\Delta_1(\vec{k})\ph^{+-}_{jl}(\vec{k})+\Delta_2(\vec{k})\ph^{--}_{jl}(\vec{k})\Big]F^{--}_{lk}(\vec{k})\\
        &+\Big[\Delta_1(\vec{k})\ph^{++}_{jl}(\vec{k})+\Delta_2(\vec{k})\ph^{-+}_{jl}(\vec{k})\Big]F^{+-}_{lk}(\vec{k})\\
        &-\Big[\Delta_1(\vec{k})\ph^{--}_{jl}(\vec{k})+\Delta_2(\vec{k})\ph^{+-}_{jl}(\vec{k})\Big]F^{-+}_{lk}(\vec{k})\Big\},
\end{aligned}
\end{equation}
where 
\begin{equation}
    \begin{gathered}
        \Delta_1(\vec{k})=D_1-D_2=\frac{\mc{V}(\Omega,\vec{K},\vec{k})}{\mc{V}^2(\Omega,\vec{K},\vec{k})+\nu_-^2\kappa^4(\vec{K},\vec{k})},\\
        \Delta_2(\vec{k})=D_3-D_4=-\frac{\ii\nu_-\kappa^2}{\mc{V}^2(\Omega,\vec{K},\vec{k})+\nu_-^2\kappa^4(\vec{K},\vec{k})},\\
                \mc{V}(\Omega,\vec{K},\vec{k})\doteq\Omega'+\ii\nu_+[k^2+(\vec{k}+\vec{K})^2],\\
        \kappa^2(\vec{K},\vec{k})\doteq (\vec{k}+\vec{K})^2-k^2.
    \end{gathered}
\end{equation}
Finally, noting that:
\begin{gather}
    \ph^{\pm\pm}_{ij}(\vec{k})=M_{ijk}(\vec{k})\pz^\mp_k,\qquad \ph_{ij}^{\pm\mp}(\vec{k})=N_{ijk}(\vec{k})\pz_k^\mp,
\end{gather}
where
\begin{gather}\label{eq:themns}
    M_{ijk}=\Big(\delta_{ij}-\frac{(k_i+K_i)K_j}{(\vec{k}+\vec{K})^2}\Big)k_k, \qquad N_{ijk}=\delta_{ik}K_j-\frac{(k_i+K_i)K_j}{(\vec{k}+\vec{K})^2}k_k,
\end{gather}
we have:
\begin{gather}
\begin{aligned}\label{eq:gv}
        G_{ij}^{(\vec{v})}(\Omega,\vec{K})=\frac{\epsilon_{imn}}{2}\int\dd\vec{k}\Big\{M_{mlj}&[\Delta_1(F_{ln}^{-+}-F_{ln}^{+-})+\Delta_2(F_{ln}^{++}-F_{ln}^{--})]\\
    +N_{mlj}&[\Delta_1(F_{ln}^{++}-F_{ln}^{--})+\Delta_2(F_{ln}^{-+}-F_{ln}^{+-})]\Big\},
\end{aligned}\\
\begin{aligned}\label{eq:gb}
        G_{ij}^{(\vec{b})}(\Omega,\vec{K})=\frac{\epsilon_{imn}}{2}\int\dd\vec{k}\Big\{M_{mlj}&[\Delta_1(F_{ln}^{-+}+F_{ln}^{+-})-\Delta_2(F_{ln}^{++}+F_{ln}^{--})]\\
    +N_{mlj}&[\Delta_1(F_{ln}^{++}+F_{ln}^{--})-\Delta_2(F_{ln}^{-+}+F_{ln}^{+-})]\Big\}.
\end{aligned}
\end{gather}

Equations \Eq{eq:gv} and \Eq{eq:gb} are the exact (within the QLA) forms of the linear nonlocal response kernel of the turbulent EMF to mean velocity and magnetic fields, respectively. We emphasise that, in deriving them, we have made no additional approximations, such as scale separation or any particular symmetries of the turbulence (\eg isotropy).
It is immediately evident from \Eq{eq:gv} that the contribution to the EMF from the mean flow vanishes if \m{\matr{F}^{++}=\matr{F}^{--}} and \m{\matr{F}^{+-}=\matr{F}^{-+}}, that is, if \m{\fl{\vec{z}}^+} and \m{\fl{\vec{z}}^-} have the same two-point correlations. This condition is somewhat trivially satisfied by hydrodynamic turbulence (in which \m{\fl{\vec{b}}=0} such that \m{\fl{\vec{z}}^+=\fl{\vec{z}}^-}) but can also be satisfied by more general MHD turbulence, as will be discussed shortly. 

In the remainder of this section, we examine \Eq{eq:gv} and \Eq{eq:gb} for specific turbulent backgrounds. We also compare them with traditional mean-field theories and the numerically inferred form of the nonlocal response kernel \m{\matr{G}^{(\vec{b})}} used in the literature.

\subsection{Hydrodynamic turbulence}\label{sec:nonlocalhydro}
Consider homogeneous hydrodynamic turbulence, where $\fl{\vec{z}}^+=\fl{\vec{z}}^-$ and thus \m{\matr{F}\doteq\matr{F}^{\pm\pm}=\matr{F}^{\pm\mp}}. Then \Eq{eq:polEMFfinal} becomes simply
\begin{equation}\label{eq:polEMFhydro}
\begin{gathered}
        \pemf_i=\epsilon_{ijk}\int\dd\vec{k}\,\Delta(\vec{k})[(\vec{\pb}\cdot\vec{k})\delta_{jl}-\pb_j K_l]F_{lk},
\end{gathered}
\end{equation}
where 
\begin{equation}
    \Delta(\vec{k})\doteq\Delta_1(\vec{k})-\Delta_2(\vec{k})=\frac{1}{\Omega'+\ii\nu k^2+\ii\eta (\vec{k}+\vec{K})^2}.
\end{equation}
This is the generic answer for hydrodynamic turbulence for general \m{\matr{F}(\vec{k})}. Notably, the EMF is entirely indepenent of \m{\av{\vec{v}}} in this case.

If we further assume that the turbulent flow is isotropic \Eq{eq:isowigner}, we have
\begin{equation}\label{eq:isohydrowigner}
    \begin{aligned}
        F_{lp}(\vec{k})=&\Big(\delta_{lp}-\frac{k_lk_p}{k^2}\Big)\frac{E(k)}{4\pi k^2}-\ii\epsilon_{lpq}\frac{k_q}{k^2}\frac{H(k)}{8\pi k^2},
    \end{aligned}
\end{equation}
where \m{E(k)} and \m{H(k)} are the energy and helicity spectra, respectively:
\begin{equation}
    \frac{1}{2}\flav{ v^2}=\int_0^\infty \dd k \,E(k),\qquad 
    \flav{ \vec{v}\cdot(\nabla\times\vec{v})}=\int_0^\infty \dd k \,H(k).
\end{equation}

We now assume sufficiently weak dissipation, such that \m{\tau_c\nu_\pm k^2\ll1} and \m{\tau_c\nu_\pm k K \ll1} over the range where \m{E(k)} and \m{H(k)} substantially contribute to the integrals over \m{\vec{k}}. In this case, we can use the following approximation for \m{\Delta}:
\begin{equation}\label{eq:deltaweakdiss}
\begin{gathered}\Delta(\vec{k})\approx\frac{1}{\Omega'+\ii\eta K^2}\Bigg(1-\frac{2\ii(\nu_+k^2+\eta \vec{k}\cdot\vec{K})}{\Omega'+\ii\eta K^2}\Bigg).
\end{gathered}
\end{equation}
Note that we have assumed dissipation to be weak, so that the terms \m{\nu_+ k^2} and \m{\eta \vec{k}\cdot\vec{K}} could be taken out of the denominator of \m{\Delta}. (The \m{\eta K^2} terms are retained for reasons that will become evident shortly.) With \Eq{eq:deltaweakdiss}, the integration over \m{\vec{k}} in \Eq{eq:polEMFhydro} can be performed without assuming any particular form of the spectra \m{E(k)} and \m{H(k)}, and one obtains the relatively familiar form:
\begin{subequations}\label{eq:polEMFweakdiss}
    \begin{gather}
        \vec{\pemf}=\mc{A}(\Omega,\vec{K})\vec{\pb}-\mc{B}(\Omega,\vec{K})(\ii\vec{K}\times\vec{\pb}),\\
        \mc{A}(\Omega,\vec{K})=\frac{\alpha}{N(\Omega,\vec{K})}+\frac{2}{3}\frac{\nu_+\tau_c^2\flav{\fl{\vec{w}}\cdot\nabla\times\fl{\vec{w}}}}{N^2(\Omega,\vec{K})},\\
         \mc{B}(\Omega,\vec{K})=\frac{\beta}{N(\Omega,\vec{K})}-\frac{2}{3}\frac{\nu_+\tau_c^2\flav{\fl{\vec{v}}\cdot\nabla\times\fl{\vec{w}}}}{N^2(\Omega,\vec{K})},
    \end{gather}
\end{subequations}
where
\begin{equation}
    N(\Omega,\vec{K})\doteq1-\ii\tau_c\Omega +\tau_c\eta K^2,
\end{equation}
\m{\alpha} and \m{\beta} are the local transport coefficients defined in \Eq{eq:localab}, 
\begin{gather}
    \flav{\vec{\fl{w}}\cdot\nabla\times\vec{\fl{w}}}=\int_0^\infty \dd k\, k^2 H(k), \qquad\flav{\vec{\fl{v}}\cdot\nabla\times\vec{\fl{w}}}=\int_0^\infty \dd k\, k^2 E(k),
\end{gather}
and the corresponding response kernel is given by
\begin{equation}\label{eq:responsekernel}
    G_{ij}^{(\vec{b})}(\Omega,\vec{K})=\mc{A}(\Omega,\vec{K})\delta_{ij}+\ii\mc{B}(\Omega,\vec{K})\epsilon_{ijk}K_k.
\end{equation}
The first terms in the expressions for \m{\mc{A}} and \m{\mc{B}} are exactly the empirical nonlocal response kernel \Eq{eq:numkernel}, with \m{\tau=\tau_c} and 
\begin{equation}\label{eq:lcalc}
    l^2=\tau_c\eta.
\end{equation}
The second terms are viscous and resistive corrections. In particular, note that in the ideal limit (\m{\nu_\pm\to 0}) and for short correlation times (\m{\tau_c\Omega\ll 1}), we recover the original local statement of the \m{\alpha}-effect, that is, \m{\mc{A}\to\alpha} and \m{\mc{B}\to\beta}.

These results are consistent with reports of the nonlocality parameter \m{\tau} being of the order of the turnover time across multiple simulations with different values of the magnetic Reynolds number. In contrast, the inferred value of \m{l} varies more widely in the literature \citep{ref:rheinhardt12,ref:brandenburg18}, and \Eq{eq:lcalc} can be considered the first actual calculation of \m{l} from first principles (modulo the fact that we rely on the QLA and introduce \m{\tau_c} in \Eq{eq:modWME} ad hoc, as usual). 

\subsection{Isotropic MHD turbulence}

The nonlocal turbulent EMF for isotropic MHD turbulence \Eq{eq:isowigner} can be written as follows:
\begin{equation}\label{eq:isomhd}
    \vec{\pemf}=\mc{A}(\Omega,\vec{K})\vec{\pb}-\mc{B}(\Omega,\vec{K})(\ii\vec{K}\times\vec{\pb})+\mc{C}(\Omega,\vec{K})\vec{\pv}-\mc{D}(\Omega,\vec{K})(\ii\vec{K}\times\vec{\pv}),\\
\end{equation}
where 
\begin{subequations}
    \begin{gather} 
        \begin{aligned}\label{eq:Anonlocal}
                    \mc{A}(\Omega,\vec{K})=\ii\int\dd \vec{k}\,\frac{k_{\vec{\pb}}^2}{k^2}\Bigg[\frac{\vec{k}\cdot\vec{K}+K^2}{(\vec{k}+\vec{K})^2}(\Delta_1+\Delta_2)\Big(\frac{H^C(k)}{8\pi k^2}- \frac{H^S(k)}{8\pi k^2}&\Big)\\
        +2\Big(\Delta_2\frac{H^S(k)}{8\pi k^2}-\Delta_1\frac{H^C(k)}{8\pi k^2}&\Big)\Bigg],
        \end{aligned}\\
        \begin{aligned}
            \mc{B}(\Omega,\vec{K})=\ii\int\dd \vec{k}\Bigg[\frac{k_{\vec{\pb}}^2}{k^2}\frac{k^2+\vec{k}\cdot\vec{K}}{(\vec{k}+\vec{K})^2}(\Delta_1+\Delta_2)\Big(\frac{E^C(k)}{4\pi k^2}- \frac{E^S(k)}{4\pi k^2}&\Big)\\
        +\frac{k^2-k_{\vec{K}}^2}{k^2}\Big(\Delta_1\frac{E^S(k)}{4\pi k^2}-\Delta_2\frac{E^C(k)}{4\pi k^2}&\Big)\Bigg],
        \end{aligned}\\
        \mc{C}(\Omega,\vec{K})=\ii\int\dd \vec{k}\,\frac{k_{\vec{\pv}}^2}{k^2}\Bigg[\frac{\vec{k}\cdot\vec{K}+K^2}{(\vec{k}+\vec{K})^2}(\Delta_1+\Delta_2)-2\Delta_2\Bigg]\frac{H^A(k)}{8\pi k^2},\label{eq:nonlocalC}\\
        \mc{D}(\Omega,\vec{K})=\ii\int\dd \vec{k}\Bigg[\frac{k_{\vec{\pv}}^2}{k^2}\frac{k^2+\vec{k}\cdot\vec{K}}{(\vec{k}+\vec{K})^2}(\Delta_1-\Delta_2)+\frac{k_{\vec{K}}^2-k^2}{k^2}\Delta_1\Bigg]\frac{E^A(k)}{4\pi k^2},
    \end{gather}
\end{subequations}
\m{k_{\vec{\xi}}\doteq(\vec{k}\cdot\vec{\xi})/\xi} denotes the component of \m{\vec{k}} along a given vector \m{\vec{\xi}} and the functions \m{E^\chi(k)} and \m{H^\chi(k)} are defined in \Eq{eq:spectra}. In particular, for ideal MHD in the commonly assumed GO limit, the above coefficients acquire a more familiar form:
\begin{subequations}\label{eq:isoalphasswd}
    \begin{gather}
                    \mc{A}(\Omega,\vec{K})\approx-\frac{\ii}{3\Omega'}\flav{\vec{\fl{v}}\cdot\nabla\times\vec{\fl{v}}-\vec{\fl{b}}\cdot\nabla\times\vec{\fl{b}}},\label{eq:Alocal}\\ 
            \mc{B}(\Omega,\vec{K})\approx\frac{\ii}{3\Omega'}\flav{\vec{\fl{v}}\cdot\vec{\fl{v}}},\\
        \mc{C}(\Omega,\vec{K})\approx 0,\\
        \mc{D}(\Omega,\vec{K})=\frac{\ii}{3\Omega'}\flav{\vec{\fl{v}}\cdot\vec{\fl{b}}}.
    \end{gather}
\end{subequations}

Let us now discuss each of these contributions to the nonlocal EMF and their relationship to the results of local mean-field theories. As established in the previous discussion for hydrodynamic turbulence (\Sec{sec:nonlocalhydro}), the transport coefficients \m{\mc{A}} and \m{\mc{B}} correspond to the usual \m{\alpha}-effect driven by kinetic helicity and turbulent diffusivity, respectively. Note that for MHD turbulence, the MFWK framework captures the cancellation of kinetic helicity by current helicity, \ie the magnetic \m{\alpha}-effect, also known as the Pouquet effect \citep{ref:pouquet76}. The coefficient \m{\mc{A}} is therefore qualitatively in agreement with previous theories, up to nonlocal corrections (the difference between \Eq{eq:Anonlocal} and \Eq{eq:Alocal}). Notably, it is commonly stated that the magnetic \m{\alpha}-effect is captured only beyond the QLA, as done in the eddy-damped quasi-normal Markovian (EDQNM) \citep{ref:orszag70} or MTA \citep{ref:blackman02} closures. However, as can be seen from our calculation, this is not the case. Although we included an MTA-like ad-hoc damping term in \Eq{eq:modWME}, this modification to the QLA is inessential for capturing the magnetic \m{\alpha}-effect in \Eq{eq:Alocal}, and taking \m{\tau_c\to\infty} results only in replacing \m{\Omega'} with \m{\Omega}. The coefficient \m{\mc{B}} also agrees with the vast majority of previous mean-field theories in that it contains no contribution to the turbulent diffusivity from magnetic fluctuations \citep{ref:vainshtein83,ref:radler02,ref:squire15_a}, although such a contribution was found with the two-scale direct interaction approximation (TSDIA) in \citep{ref:yoshizawa90}.

The contribution to the turbulent EMF proportional to the mean-flow vorticity, as determined by \m{\mc{D}}, is known as the Yoshizawa effect \citep{ref:yoshizawa90,ref:yokoi13} and is related to the cross-helicity \m{\flav{\fl{\vec{v}}\cdot\fl{\vec{b}}}}. As with the \m{\alpha}-effect term, the coefficient \m{\mc{D}} agrees exactly with previous mean-field theories in the GO limit (\m{lK\ll 1}) and for short correlation times (\m{\tau_c\Omega\ll1}). 

In contrast, the effect of \m{\mc{C}} predicted by the MFWK framework is qualitatively new. The term in the EMF that is proportional to the mean flow has received limited attention in the literature. Typically, Galilean invariance is invoked to argue that any term proportional to the mean flow should vanish. This is, of course, true if the flow is homogeneous, and, indeed, the contribution from \m{\av{\vec{v}}} vanishes in the limit of negligible \m{K}; i.e. \m{\mathcal{C} \to 0} as \m{K\to0}. However, the contribution of \m{\av{\vec{v}}} at nonnegligible \m{K} is nonzero.
To the best of our knowledge, the only other work where this subject is discussed is \citep{ref:radler10}. Using our notation, the result of \citep{ref:radler10} can be expressed as follows:
\begin{subequations}\label{eq:radlerresult}
    \begin{gather}
\emf=\hdots+C\vec{\av{\vec{v}}},\\
C=\frac{1}{3}\int\dd \tau\int \dd \vec{\xi} \,[\mc{G}^{(\nu)}(\tau,\xi)-\mc{G}^{(\eta)}(\tau,\xi)]\flav{\fl{\vec{v}}(t,\vec{x})\cdot\fl{\vec{j}}(t-\tau,\vec{x}-\vec{\xi})},\\
\mc{G}^{(\gamma)}(\tau,\xi)\doteq\begin{cases}
(4\pi\gamma\tau)^{-3/2}\exp(-\xi^2/4\gamma\tau),& \tau>0,\\
0,&\tau\leq0.
\end{cases}
\end{gather}
\end{subequations}
The ellipsis indicates the additional contributions to the EMF that are not relevant to the present discussion.

Although \Eq{eq:radlerresult} and \Eq{eq:nonlocalC} are in agreement in that the basic effect is driven by flow current alignment \m{\flav{\fl{\vec{v}}\cdot\fl{\vec{j}}}}, they differ in several significant ways beyond what can be attributed to nonlocal corrections. Firstly, \Eq{eq:radlerresult} predicts a finite contribution to the nonlocal EMF even for constant mean flows. In \citep{ref:radler10}, it is argued that this does not violate Galilean invariance because the frame of reference is fixed by the assumption of isotropic turbulence, \ie that shifting from a frame with a constant mean flow and isotropic turbulence to the frame where the mean flow is zero would render the turbulence anisotropic. However, because the constant shift to the total velocity \m{\vec{v}=\av{\vec{v}}+\fl{\vec{v}}} that would be introduced by a change in reference frame would be absorbed by the mean flow \m{\av{\vec{v}}} and not contribute to the fluctuating fields \m{\fl{\vec{v}}} that comprise the turbulence. In contrast, \Eq{eq:nonlocalC} has the correct limiting behaviour for constant mean flows, as the transport coefficient \m{\mc{C}} vanishes as \m{K\to 0}.

Although the violation of Galilean invariance alone indicates that \Eq{eq:radlerresult} is unphysical, let us also mention other ways in which it disagrees with our theory. Equation \Eq{eq:radlerresult} predicts that the mean-flow contribution to the EMF vanishes when \m{\nu=\eta}, and, in particular, in the ideal limit \m{\nu_\pm\to 0}. In contrast, \Eq{eq:nonlocalC} at  \m{\nu=\eta} (\ie \m{\nu_-=0}) predicts nonzero \m{C}:
\begin{equation}
            \mc{C}(\Omega,\vec{K})=\ii\int\dd \vec{k}\,\frac{k_{\vec{\pv}}^2}{k^2}\frac{\vec{k}\cdot\vec{K}+K^2}{(\vec{k}+\vec{K})^2[\Omega'+\ii\nu_+(k^2+(\vec{k}+\vec{K})^2]}\frac{H^A(k)}{8\pi k^2},
\end{equation}
or, in the ideal limit,
\begin{equation}
    \mc{C}(\Omega,\vec{K})=\frac{\ii}{\Omega'}\int\dd \vec{k}\,\frac{k_{\vec{\pv}}^2}{k^2}\frac{\vec{k}\cdot\vec{K}+K^2}{(\vec{k}+\vec{K})^2}\frac{H^A(k)}{8\pi k^2},\\
\end{equation}
both of which are generally nonzero. These discrepancies may be related to the treatment of the mean flow as an externally prescribed field, as opposed to a self-consistent field.
 
In the following section, we will see that this little-known contribution to the EMF couples the mean magnetic fields and flows in such a way that enables a previously unknown \m{\flav{\fl{\vec{v}}\cdot\fl{\vec{j}}}}-driven dynamo effect.

\section{Modulational modes of MHD turbulence}\label{sec:modmodes}
Perhaps some of the most basic questions that can be asked regarding mean-field generation from turbulence are: what properties of turbulence make it susceptible to spontaneously generating mean fields, what do these unstable mean fields look like, and how fast do they grow? 
These questions lie within the purview of the modulational-mode analysis presented in this section, which therefore encapsulates some of the most fundamental predictions of MFWK as a theoretical framework. We will also compare and contrast the predictions of MFWK with those of existing mean-field theories.
\subsection{Derivation}
The dispersion relation of modulational modes follows directly from the linearised MFWK equations written in terms of the polarizations of the perturbed quantities \Eq{eq:polmf} and \Eq{eq:polWME}. 
Let us first rewrite \Eq{eq:polmf} in a form that aids the eventual dispersion matrix formulation:
\begin{equation}\label{eq:predispmat}
    \begin{pmatrix}
        (\ii\Omega- \nu_+K^2) \matr{1}_3&-\nu_- K^2 \matr{1}_3\\
        -\nu_- K^2 \matr{1}_3&(\ii\Omega- \nu_+K^2) \matr{1}_3
    \end{pmatrix}\vec{\pz}=\ii\begin{pmatrix}
        \vec{\mc{S}}^+\\\vec{\mc{S}}^-
    \end{pmatrix},
\end{equation}
where \m{\vec{\pz}\doteq(\vec{\pz}^+,\vec{\pz}^-)^\intercal} and 
\begin{equation}\label{eq:sourcepol}
\begin{aligned}
    \mc{S}^\pm_i&=\int\dd\vec{k}\,K_n\Big(\pf_{in}^{\pm\mp}-\frac{K_iK_m}{K^2}\pf_{mn}^{\pm\mp}\Big)\\
    &=K_n\Big(\delta_{im}-\frac{K_iK_m}{K^2}\Big)\int\dd\vec{k}\,\pf_{mn}^{\pm\mp}.
\end{aligned}
\end{equation}
To find the dispersion matrix, we must express \m{\vec{\mc{S}}} in terms of the mean fields \m{\vec{\pz}^\pm}. To do this, the integral of the perturbed Wigner matrix \Eq{eq:offdiagf} can be simplified by shifting and flipping integration variables as done in \Sec{sec:nonlocalemfderiv}:
\begin{equation}
    \begin{aligned}
        \int\dd\vec{k}\,\pf_{mn}^{\pm\mp}=\Bigg[\int\dd\vec{k}\, M_{mpj}(D_2F_{pn}^{\mp\pm}+D_3F_{pn}^{\mp\mp})+N_{mpj}(D_2F_{pn}^{\pm\pm}+D_3F_{pn}^{\pm\mp})&\\
        +M_{npj}(D_1F_{pm}^{\mp\pm}+D_4F_{pm}^{\mp\mp})+N_{npj}(D_1F_{pm}^{\pm\pm}+D_4F_{pm}^{\pm\mp})&\Bigg]\pz_j^\pm\\
        +\Bigg[\int\dd\vec{k}\, M_{mpj}(D_1F_{pn}^{\pm\mp}+D_4F_{pn}^{\pm\pm})+N_{mpj}(D_1F_{pn}^{\mp\mp}+D_4F_{pn}^{\mp\pm})&\\
        +M_{npj}(D_2F_{pm}^{\pm\mp}+D_3F_{pm}^{\pm\pm})+N_{npj}(D_2F_{pm}^{\mp\mp}+D_3F_{pm}^{\mp\pm})&\Bigg]\pz_j^\mp,
    \end{aligned}
\end{equation}
where the matrices \m{M_{ijk}} and \m{N_{ijk}} are defined in \Eq{eq:themns} and the functions \m{D_n} are defined as follows:
\begin{equation}
    \begin{gathered}
        D_1=\frac{\mc{W}(\mc{W}^2+\nu_-^2\kappa_1^4)}{(\mc{W}^2+\nu_-^2\kappa_1^4)^2-4\nu_-^4\kappa_2^8},\\D_2=\frac{-2\nu_-^2\kappa_2^4\mc{W}}{(\mc{W}^2+\nu_-^2\kappa_1^4)^2-4\nu_-^4\kappa_2^8},\\
                D_3=\frac{-\ii\nu_-(\vec{k}+\vec{K})^2(\mc{W}^2+\nu_-^2\kappa_1^4)+2\ii\nu_-^3k^2\kappa_2^4}{(\mc{W}^2+\nu_-^2\kappa_1^4)^2-4\nu_-^4\kappa_2^8},\\D_4=\frac{-\ii\nu_-k^2(\mc{W}^2+\nu_-^2\kappa_1^4)+2\ii\nu_-^3(\vec{k}+\vec{K})^2\kappa_2^4}{(\mc{W}^2+\nu_-^2\kappa_1^4)^2-4\nu_-^4\kappa_2^8},\\
    \end{gathered}
\end{equation}
and \m{\mc{W}\doteq\Omega'+\ii\nu_+\kappa_0^2}, \m{\kappa_0^2\doteq k^2+(\vec{k}+\vec{K})^2}, \m{\kappa_1^4\doteq k^4+(\vec{k}+\vec{K})^4}, and \m{\kappa_2^4\doteq k^2(\vec{k}+\vec{K})^2}.

After extensive but straightforward calculation, one can rewrite this as follows:

\begin{equation}
    \begin{gathered}
        \ii\matr{S}^\pm=\matr{P}^\pm \vec{\pz}^\pm+\matr{Q}^\pm\vec{\pz}^\mp,
    \end{gathered}
\end{equation}
where the matrices \m{\matr{P}^\pm} and \m{\matr{Q}^\pm} are given by
\begin{subequations}
    \begin{gather}
        \begin{aligned}\label{eq:pij}            
        P_{ij}^\pm=\ii K_n\int\dd\vec{k}\Big[\,&\frac{K_p}{K^2}F_{pn}^{\mp\mp}\mc{M}_{ij}(D_{12},D_1,D_{34},D_{13})-F^{\mp\mp}_{in}D_3k_j\\
        +&\frac{K_p}{K^2}F_{pn}^{\mp\pm}\mc{M}_{ij}(D_{34},D_4,D_{12},D_{24})-F^{\mp\pm}_{in}D_2k_j\\
        -&F^{\mp\mp}_{ni}\mc{N}_j(D_2,D_4,D_{24})-F^{\mp\pm}_{ni}\mc{N}_j(D_3,D_1,D_{13})\Big],
        \end{aligned}\\
        \begin{aligned}\label{eq:qij}
        Q^\pm_{ij}=\ii K_n\int\dd\vec{k}\Big[\,&\frac{K_p}{K^2}F_{pn}^{\pm\pm}\mc{M}_{ij}(D_{12},D_2,D_{34},D_{24})-F^{\pm\pm}_{in}D_4k_j\\
        +&\frac{K_p}{K^2}F_{pn}^{\pm\mp}\mc{M}_{ij}(D_{34},D_3,D_{12},D_{13})-F^{\pm\mp}_{in}D_1k_j\\
        -&F^{\pm\pm}_{ni}\mc{N}_j(D_1,D_3,D_{13})-F^{\pm\mp}_{ni}\mc{N}_j(D_4,D_2,D_{24})\Big],
        \end{aligned}
    \end{gather}
\end{subequations}
where
\begin{subequations}
    
\end{subequations}
\begin{gather}
\begin{aligned}
        &\mc{M}_{ij}(f_1,f_2,f_3,f_4)\doteq\\&f_1 K_iK_j-f_2K^2\delta_{ij}+\Big(f_3-D\frac{\vec{K}\cdot(\vec{k}+\vec{K})}{(\vec{k}+\vec{K})^2}\Big)K_ik_j+f_4\frac{K^2}{(\vec{k}+\vec{K})^2}(k_i+K_i)K_j,\\
\end{aligned}\\
\mc{N}_{j}(f_1,f_2,f_3)\doteq f_1K_j+\Big(f_2-f_3\frac{\vec{K}\cdot(\vec{k}+\vec{K})}{(\vec{k}+\vec{K})^2}\Big)k_j,
\end{gather}
and we use the shorthand \m{D\doteq D_1+D_2+D_3+D_4} and \m{D_{mn}\doteq D_m+D_n}.

\subsection{Dispersion relation}
The general dispersion relation and polarizations for the modulational modes of MHD turbulence are given by
\begin{gather}\label{eq:gendisprel}
            \det \matr{\Pi}=0, \qquad    \matr{\Pi}\vec{\pz}=0,
\end{gather}
where \m{\vec{\pz}\doteq(\vec{\pz}^+,\vec{\pz}^-)^\intercal}, \m{\av{\vec{z}}^\pm=\vec{\pz}^\pm\exp{(-\ii\Omega t+\ii\vec{K}\cdot\vec{x})}}, and the dispersion matrix \m{\matr{\Pi}} is given by
\begin{equation}
      \matr{\Pi}=\begin{pmatrix}
            \matr{\Pi}^{++}&\matr{\Pi}^{+-}\\
            \matr{\Pi}^{-+}&\matr{\Pi}^{--}
        \end{pmatrix},
\end{equation}
where
\begin{subequations}
    \begin{gather}
        \matr{\Pi}^{\pm\pm}=(\ii\Omega-\nu_+K^2)\matr{1}_3-\matr{P}^\pm,\\
        \matr{\Pi}^{\pm\mp}=-\nu_-K^2\matr{1}_3-\matr{Q}^\pm,
    \end{gather}
 \end{subequations}
and \m{\matr{P}^\pm} and \m{\matr{Q}^\pm} are defined in \Eq{eq:pij} and \Eq{eq:qij}, respectively. Equation \Eq{eq:gendisprel} is the general dispersion relation of modulational modes of generic incompressible resistive MHD turbulence within the MTA closure (which reduces to the QLA in the limit \m{\tau_c\to \infty}). We emphasise that no particular properties of the turbulence, \eg isotropy or scale-separation, have been assumed to derive these equations. 

\subsection{Examples}
Although \Eq{eq:gendisprel} is powerful in its generality, the intricate dependence of the integrands in \Eq{eq:pij} and \Eq{eq:qij} on \m{\Omega} and \m{\vec{K}} means that obtaining the solutions to \Eq{eq:gendisprel} is challenging in its own right. In the remainder of this section, we make a series of simplifying assumptions to obtain explicit solutions of \Eq{eq:gendisprel} that allow us to compare the predictions of MFWK to those of traditional mean-field theories.

\subsubsection{Isotropic turbulence}
For isotropic turbulent backgrounds \Eq{eq:isowigner}, we can, without loss of generality, let the modulational wavevector \m{\vec{K}} lie along \m{\uvec{z}} (\m{\vec{K}=K\uvec{z}}). By incompressibility, we then have \m{\pz_z^\pm=0}, so we may consider the following reduced system:
\begin{equation}
    \sub{\matr{\Pi}}\sub{\vec{\pol{z}}}=0,\qquad\det\sub{\matr{\Pi}}=0,
\end{equation}
where \m{\sub{\vec{\pol{z}}}\doteq(\pol{z}_x^+,\pol{z}_y^+,\pol{z}_x^-,\pol{z}_y^-}) and the dispersion matrix \m{\sub{\matr{\Pi}}} can be written as
\begin{gather}\label{eq:isodispmat}
        \sub{\matr{\Pi}}\doteq\begin{pmatrix}
    M_S+M_A&m_S+m_A&N_S+N_A&n_S+n_A\\
    -m_S-m_A&M_S+M_A&-n_S-n_A&N_S+N_A\\
    N_S-N_A&n_S-n_A&M_S-M_A&m_S-m_A\\
    -n_S+n_A&N_S-N_A&-m_S+m_A&M_S-M_A\\
            \end{pmatrix}.
            \end{gather}
Here,
\begin{subequations}
            \begin{gather}
            \label{eq:elements}
            \begin{aligned}
             M_S\doteq\ii\Omega-\nu_+K^2+\ii K\int\dd\vec{k}\frac{k_x^2}{k^2}\Big[\,\frac{E_S}{4\pi k^2}\sub{\mc{M}}(D_1,D_{13},D_{34},D_{24})&\\+\frac{E_C}{4\pi k^2}\sub{\mc{M}}(D_4,D_{24},D_{12},D_{13})&\Big],
            \end{aligned}\\
            \begin{aligned}
                           N_S\doteq-\nu_-K^2+\ii K\int\dd\vec{k}\frac{k_x^2}{k^2}\Big[\,\frac{E_S}{4\pi k^2}\sub{\mc{M}}(D_2,D_{24},D_{34},D_{13})&\\+\frac{E_C}{4\pi k^2}\sub{\mc{M}}(D_3,D_{13},D_{12},D_{24})&\Big],
            \end{aligned}\\
            M_A\doteq-\ii K\int\dd\vec{k}\frac{E_A}{4\pi k^2}\frac{k_x^2}{k^2}\sub{\mc{M}}(D_1,D_{13},D_{34},D_{24}),\\
            N_A\doteq\ii K\int\dd\vec{k}\frac{E_A}{4\pi k^2}\frac{k_x^2}{k^2}\sub{\mc{M}}(D_2,D_{24},D_{34},D_{13}),\\
            m_S\doteq K\int\dd \vec{k}\frac{k_x^2}{k^2}\Big[\frac{H_S}{8\pi k^2}\sub{\mc{N}}(D_3,D_4,D_{24})+\frac{H_C}{8\pi k^2}\sub{\mc{N}}(D_2,D_1,D_{13})],\\
            n_S\doteq K\int\dd \vec{k}\frac{k_x^2}{k^2}\Big[\frac{H_S}{8\pi k^2}\sub{\mc{N}}(D_4,D_3,D_{13})+\frac{H_C}{8\pi k^2}\sub{\mc{N}}(D_1,D_2,D_{24})\Big],\displaybreak\\
            m_A\doteq -K\int\dd \vec{k}\frac{k_x^2}{k^2}\frac{H_A}{8\pi k^2}\sub{\mc{N}}(D_3,D_4,D_{24}),\\
            n_A \doteq K\int\dd \vec{k}\frac{k_x^2}{k^2}\frac{H_A}{8\pi k^2}\sub{\mc{N}}(D_4,D_3,D_{13}),
\end{gather}
\end{subequations}
where
\begin{gather}
    \sub{\mc{M}}(f_1,f_2,f_3,f_4)\doteq2K\Big(-f_1+f_2\frac{k_x^2}{(\vec{k}+\vec{K})^2}\Big)+k_z\Big(f_3-f_4\frac{K(k_z+K)}{(\vec{k}+\vec{K})^2}\Big),\\
    \sub{\mc{N}}(f_1,f_2,f_3)\doteq f_1-\Big(f_2-f_3\frac{K(k_z+K)}{(\vec{k}+\vec{K})^2}\Big).
\end{gather}
The dispersion relation can then be factored as follows:
\begin{equation}\label{eq:subdisprel}
\begin{gathered}
        \det\sub{\matr{\Pi}}=(a+\ii b)(a-\ii b),\\
            a\doteq M_S^2-M_A^2-N_S^2+N_A^2-m_S^2+m_A^2+n_S^2-n_A^2,\\
        b\doteq2(M_sm_s-N_sn_s-M_Am_A+N_An_A).
\end{gathered}
\end{equation}
In the remainder of the section, we focus on solving \Eq{eq:subdisprel} for isotropic backgrounds as an example.

\subsubsection{Ideal limit}
The dispersion matrix \Eq{eq:isodispmat} takes a particularly simple form in the ideal limit \m{\nu_\pm\to0}. In this limit, one has \m{D_1\to 1/\Omega'} and all of the \m{D_{n\neq1}} are zero, so the modulational frequency \m{\Omega} can be pulled outside of the integrals in \Eq{eq:elements}. Then, we have 
\begin{subequations}\label{eq:ideallim}
    \begin{gather}
    \begin{aligned}
        M_S=\ii\Omega+\frac{\ii K}{\Omega'}\int\dd\vec{k}\Bigg\{\frac{E_S}{4\pi k^2}\frac{k_x^2}{k^2}\Bigg[2K\Big(-1+\frac{k_x^2}{(\vec{k}+\vec{K})^2}\Big)\Bigg]&\\
                            -\frac{E_C}{4\pi k^2}\frac{k_x^2}{k^2}\Bigg[k_z\Big(\frac{K (k_z+K)}{(\vec{k}+\vec{K})^2}\Big)\Bigg]&\Bigg\},
    \end{aligned}\\
    \begin{aligned}
        N_S=\frac{\ii K}{\Omega'}\int\dd\vec{k}\Bigg\{\frac{E_S}{4\pi k^2}\frac{k_x^2}{k^2}\Bigg[k_z\Big(-\frac{K(k_z+K)}{(\vec{k}+\vec{K})^2}\Big)\Bigg]&\\+\frac{E_C}{4\pi k^2}\frac{k_x^2}{k^2}\Bigg[2K\Big(\frac{k_x^2}{(\vec{k}+\vec{K})^2}\Big)\Bigg]&\Bigg\},
    \end{aligned}\\
            M_A=-\frac{\ii K}{\Omega'}\int\dd\vec{k}\frac{E_A}{4\pi k^2}\frac{k_x^2}{k^2}\Bigg[2K\Big(-1+\frac{k_x^2}{(\vec{k}+\vec{K})^2}\Big)\Bigg],\\
            N_A=\frac{\ii K}{\Omega'}\int\dd\vec{k}\frac{E_A}{4\pi k^2}\frac{k_x^2}{k^2}\Bigg[k_z\Big(-\frac{K(k_z+K)}{(\vec{k}+\vec{K})^2}\Big)\Bigg],\\
            m_S= \frac{K}{\Omega'}\int\dd \vec{k}\frac{k_x^2}{k^2}\Bigg\{\frac{H_C}{8\pi k^2}\Bigg[-1+\frac{K(k_z+K)}{(\vec{k}+\vec{K})^2}\Bigg]\Bigg\},\\
            n_S=\frac{K}{\Omega'}\int\dd \vec{k}\frac{k_x^2}{k^2}\Bigg\{\frac{H_S}{8\pi k^2}\Bigg[\frac{K(k_z+K)}{(\vec{k}+\vec{K})^2}\Bigg]+\frac{H_C}{8\pi k^2}\Bigg\},\\
            m_A=0,\\
            n_A =\frac{K}{\Omega'}\int\dd \vec{k}\frac{k_x^2}{k^2}\frac{H_A}{8\pi k^2}\Bigg[\frac{K(k_z+K)}{(\vec{k}+\vec{K})^2}\Bigg].
    \end{gather}
\end{subequations}
Hence, the dispersion relation \Eq{eq:subdisprel} becomes the product of two quadratic polynomials in \m{\Omega\Omega'}. It can then be easily solved, yielding eight total solutions \m{\Omega(\vec{K})}. The solutions obtained in this manner are what will be presented throughout this section. Also, unless explicitly stated otherwise, we will assume the limit \m{\St\to\infty}, \ie \m{\Omega'\approx \Omega}, for clarity.

Although the \m{\Rm,\,\St\to\infty} limit is formally outside of the QLA applicability domain, we believe this to be a worthwhile approach nonetheless, for the following reasons. Firstly, it is often the case that equations derived in a certain asymptotic limit qualitatively hold outside of that same limit. In particular, the accuracy of the QLA for the modulational dynamics of MHD is a rather complicated issue, and the conventional validity criterion min\m{(\St,\Rm)\ll 1} can be understood as a sufficient, but not necessary condition \citep{ref:jin24}. Secondly, the simplified picture that emerges in the QLA can serve as a stepping stone towards a more complex future theory that will contain quasilinear dynamics as a limiting case. 

In addition, we will also supplement qualitatively novel predictions of MFWK with a study of their dependence on \m{\St} within the MTA-like closure used in \Eq{eq:modWME}, \ie retaining the \m{\tau_c^{-1}} term in \m{\Omega'=\Omega+\ii\tau_c^{-1}}. This is, of course, a highly simplified model of the effect of the higher-order correlations neglected in the QLA, the results of which must therefore be taken with a grain of salt. Nevertheless, establishing the dependence on \m{\St} within this limited approach can serve as a tentative predictor of the robustness of the collective effects found in the ideal limit (\m{\St\gg1}) to eddy--eddy interactions.
\subsection{Parametrization of isotropic helical background}
As discussed in \Sec{sec:equilibria}, the Wigner matrix of a generic isotropic turbulent background is fully determined by the six spectra that characterise the total energy \m{E_S(k)}, residual energy \m{E_C(k)}, cross-helicity \m{E_A(k)}, total (kinetic \m{+} current) helicity \m{H_S(k)}, residual (kinetic \m{-} current) helicity \m{H_C(k)}, and flow--current alignment \m{H_A(k)}. Solving \Eq{eq:subdisprel} for different background spectra will therefore allow us to determine how the modulational dynamics of the system depend on the properties of the turbulent equilibrium.

For simplicity, we assume Gaussian test spectra:
\begin{subequations}
    \begin{gather}
        \frac{E_S}{4\pi k^2}=\frac{\flav{\fl{\vec{v}}\cdot\fl{\vec{v}}}+\flav{\fl{\vec{b}}\cdot\fl{\vec{b}}}}{(\sqrt{2\pi})^3}\frac{l^3}{2}\exp\Big(-\frac{l^2 k^2}{2}\Big),\\
        \frac{E_C}{4\pi k^2}=\frac{\flav{\fl{\vec{v}}\cdot\fl{\vec{v}}}-\flav{\fl{\vec{b}}\cdot\fl{\vec{b}}}}{(\sqrt{2\pi})^3}\frac{l^3}{2}\exp\Big(-\frac{l^2 k^2}{2}\Big),\\
        \frac{E_A}{4\pi k^2}=\frac{2\flav{\fl{\vec{v}}\cdot\fl{\vec{b}}}}{(\sqrt{2\pi})^3}\frac{l^3}{2}\exp\Big(-\frac{l^2 k^2}{2}\Big),\\
                \frac{H_S}{8\pi k^2}=\frac{\flav{\fl{\vec{v}}\cdot\nabla\times\fl{\vec{v}}}+\flav{\fl{\vec{b}}\cdot\nabla\times\fl{\vec{b}}}}{(\sqrt{2\pi})^3}\frac{l^3}{2}\exp\Big(-\frac{l^2 k^2}{2}\Big),\\
        \frac{H_C}{8\pi k^2}=\frac{\flav{\fl{\vec{v}}\cdot\nabla\times\fl{\vec{v}}}-\flav{\fl{\vec{b}}\cdot\nabla\times\fl{\vec{b}}}}{(\sqrt{2\pi})^3}\frac{l^3}{2}\exp\Big(-\frac{l^2 k^2}{2}\Big),\\
        \frac{H_A}{8\pi k^2}=\frac{\flav{\fl{\vec{v}}\cdot\nabla\times\fl{\vec{b}}}+\flav{\fl{\vec{b}}\cdot\nabla\times\fl{\vec{v}}}}{(\sqrt{2\pi})^3}\frac{l^3}{2}\exp\Big(-\frac{l^2 k^2}{2}\Big).
    \end{gather}
\end{subequations}
Such spectra correspond to two-point correlations of the form
\begin{equation}
    \flav{\fl{\vec{v}}(\vec{x}+\vec{r}/2)\cdot\fl{\vec{v}}(\vec{x}-\vec{r}/2)}=\flav{\fl{\vec{v}}\cdot\fl{\vec{v}}}\exp\Big(-\frac{r^2}{2l^2}\Big)
\end{equation}
and are a convenient and popular choice in the mean-field dynamo literature \citep{ref:radler06,ref:squire15_a}. Note that we have assumed the same Gaussian form and same characteristic correlation length \m{l} for all spectra. Although this is surely a simplification, it enables a minimal parametrization of the properties of the turbulent background in terms of the single-point statistics. Although six spectra are needed to fully define the isotropic Wigner matrix, we need only five parameters to capture physical properties of the turbulent equilibria that have the potential to qualitatively impact the modulational dynamics, since 
\begin{equation}
    \tau\doteq l/v_\rms
\end{equation}
simply serves to set the characteristic time scale of the problem. In the following section, we will normalise all modulational frequencies, \m{\Omega} to \m{\tau^{-1}} and all modulational wavevectors \m{\vec{K}} to \m{l^{-1}}. Then, our turbulent equilibria (after the Gaussian spectral ansatz) can be fully characterised by the five dimensionless parameters:
\begin{subequations}
\begin{gather}
m \doteq b_{\rms}^2/v_{\rms}^2,\\
a\doteq\flav{\fl{\vec{v}}\cdot\fl{\vec{b}}}/v_{\rms}b_{\rms},\\
        h_v\doteq l\flav{\fl{\vec{v}}\cdot\nabla\times\fl{\vec{v}}}/v_{\rms}^2,\\
        h_b\doteq l\flav{\fl{\vec{b}}\cdot\nabla\times\fl{\vec{b}}}/b_{\rms}^2,\\
        h_c\doteq l\flav{\fl{\vec{v}}\cdot\nabla\times\fl{\vec{b}}}/v_{\rms}b_{\rms},
\end{gather}
\end{subequations}
where \m{
\flav{\fl{\vec{v}}\cdot\fl{\vec{v}}}=v_{\rms}^2} and \m{\flav{\fl{\vec{b}}\cdot\fl{\vec{b}}}=b_{\rms}^2.}
The relative strength of magnetic fluctuations is captured by the parameter \m{m}, with \m{m=0} corresponding to hydrodynamic turbulence and \m{m=1} corresponding to equipartition. The parameter \m{a} is the normalised cross-helicity, which is nonzero only in the case of `imbalanced' MHD turbulence, \ie when energy is unevenly distributed between \m{\fl{\vec{z}}^+} and \m{\fl{\vec{z}}^-}. (Note that \m{\flav{\fl{\vec{z}}^+\cdot\fl{\vec{z}}^+}-\flav{\fl{\vec{z}}^-\cdot\fl{\vec{z}}^-}=2\flav{\fl{\vec{v}}\cdot\fl{\vec{b}}}}.) The parameters \m{h_v} and \m{h_b} are the normalised kinetic and current helicities, respectively. The parameter \m{h_c} captures flow--current alignment and can be understood as the `helical counterpart' to the cross-helicity \m{a} in that it captures the imbalance of the Els\"asser-fields helicities (\m{\flav{\fl{\vec{z}}^+\cdot\nabla\times\fl{\vec{z}}^+}-\flav{\fl{\vec{z}}^+\cdot\nabla\times\fl{\vec{z}}^+}=2\flav{\fl{\vec{v}}\cdot\fl{\vec{j}}}}). 

Note that the parameters are scaled such that their values lie between zero and one for `realistic' turbulence, which we define to include turbulence that can be feasibly obtained in numerical simulations with the appropriate choice of forcing functions. For example, \m{h_v=1} is commonly referred to as `maximally helical turbulence' in the literature, and, understandably, it is a popular scenario for numerical studies of the \m{\alpha}-effect \citep{ref:brandeburg01,ref:sur08,ref:mitra09}, although such turbulence is indeed highly stylised and not realistic in the usual sense of the word.

The mean-field effects associated with \m{h_v}, \m{h_b}, and \m{m} are relatively well understood, since the kinetic-helicity-driven \m{\alpha}-effect and its quenching by current helicity are arguably the central results of traditional mean-field theory and have been the subject of much numerical investigation. Furthermore, these effects alone can be accomodated within the usual mean-field treatment where only the dynamics of the mean magnetic field are considered, since the mean induction equation decouples from the momentum equation in the absence of cross-helicity or flow--current alignment, as discussed in \Sec{sec:nonlocalemf}. 

For cases where \m{a=0} and \m{h_c=0}, we therefore expect MFWK to largely corroborate the results of previous mean-field theories. In contrast, the mean-field dynamics associated with cross-helicity (as parametrised by \m{a}) and flow--current alignment (parametrised by \m{h_c}) have not yet been fully analysed. As discussed in \Sec{sec:nonlocalemf}, cross-helicity and flow--current alignment couple the mean induction and momentum equations such that the usual kinematic approach cannot be applied. Determining the mean-field modes associated with these effects has therefore, until now, been intractable in the general case.\footnote{However, see \citep{ref:courvoisier10,ref:courvoisier10b} for a different approach to this problem for simple cases.} In particular, we report a new dynamo effect driven by correlations between the fluctuating velocity and current, \m{\flav{\fl{\vec{v}}\cdot\fl{\vec{j}}}}.
\subsection{Benchmark example: the $\alpha^2$-dynamo}
We first benchmark the modulational dynamics obtained with the MFWK formalism by reproducing the expected \m{\alpha}-effect and associated \m{\alpha^2}-dynamo, \ie solve \Eq{eq:subdisprel} for \m{a=h_c=0}.
\subsubsection{Hydrodynamic turbulence}

\begin{figure}
 \centering
 \includegraphics[width=.9\textwidth]{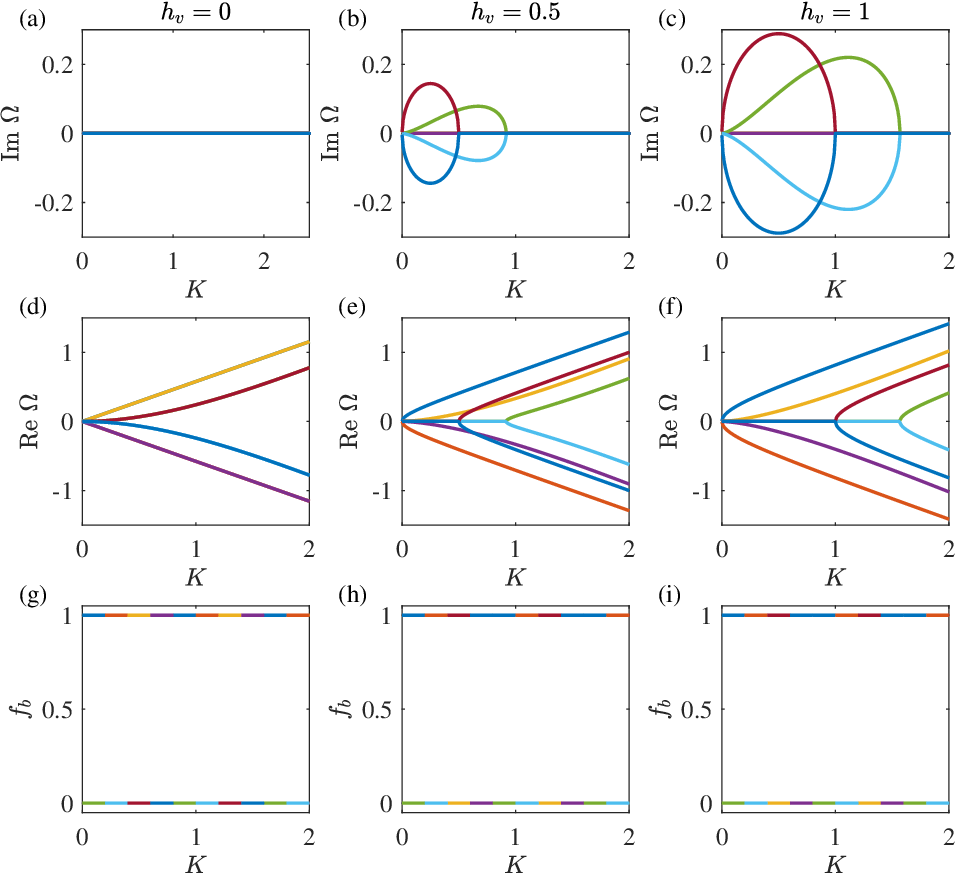}
 \caption{\label{fig:alphareference}
 The imaginary ((a)-(c)) and real ((d)-(f)) parts of modulational frequency \m{\Omega} vs. modulational wavevector \m{K} at \m{m=a=h_b=h_c=0} and \m{\St=10^5} for \m{h_v=0} ((a), (d)); \m{h_v=0.5} ((b), (e)); and \m{h_v=1} ((c), (f)). The magnetic-energy fraction \m{f_b\doteq \pb^2/(\pv^2+\pb^2)} of the corresponding eigenmodes are given in panels (g)-(i). The frequencies are given in units of the inverse turnover time \m{\tau^{-1}} and wavevectors are given in units of the inverse characteristic eddy size \m{l^{-1}}.
 }
\end{figure}

\begin{figure}
 \centering
 \includegraphics[width=.425\textwidth]{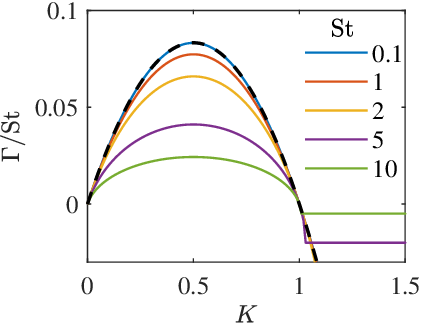}
 \caption{\label{fig:alphacomp}
The growth rate \m{\Gamma\doteq  \mathrm{ Im }\, \Omega} vs. modulational wavenumber \m{K} for various \m{\St\doteq\tau_c/\tau}. The classic \m{\alpha^2}-dynamo dispersion relation \Eq{eq:alpha2classic} is shown in the black dashed line. The frequencies are given in units of the inverse turnover time \m{\tau^{-1}} and wavevectors are given in units of the inverse characteristic eddy size \m{l^{-1}}.
 }
\end{figure}

Figure \ref{fig:alphareference} shows the modulational modes obtained from MFWK for hydrodynamic helical turbulence. As expected, the kinetic helicity supports a stationary (\m{\re \Omega =0}) unstable solution that is helically \m{\vec{\pb}}-polarised (\m{\pb_x=\ii\pb_y}, \m{\pv_x=\pv_y=0}), with a growth rate that is maximised at relatively large modulational wavelengths (\m{lK<1}). To further confirm that the unstable \m{\vec{\pb}}-polarised mode is indeed the expected \m{\alpha^2}-dynamo, \Fig{fig:alphacomp} shows that the growth rate of the unstable \m{\vec{\pb}}-polarised mode indeed converges to the local \m{\alpha^2}-dynamo growth rate,
\begin{equation}\label{eq:alpha2classic}
    \Gamma=\alpha K-(\eta+\beta) K^2,
\end{equation}
in the \m{\St\ll 1} limit, and still qualitatively agrees with \Eq{eq:alpha2classic} for larger values of \m{\St} in that it is unstable strictly for \m{K<1} and is stabilised by current helicity (see \Fig{fig:dynamocomp}).

Note that in the ideal limit (\m{\nu_\pm\to0}), and unless driven unstable by helicity, the modulational perturbations obtained from the MFWK framework are damped at the rate \m{\tau_c^{-1}/2}, irrespective of the modulational wavenumber \m{K}. In this sense, the interpretation of \m{\beta} in the local expansion of the EMF (also, \m{\mc{B}} in \Eq{eq:isomhd}) as a turbulent diffusivity relies on the assumption of short correlation times, and the analogy must therefore be applied with caution. A true dissipation such as viscosity \m{\nu} would damp the modulational perturbations at the rate \m{\nu K^2}.  

Figure \ref{fig:alphareference} also shows another (less) unstable stationary mode that is helically \m{\vec{\pv}}-polarised (\m{\ii\pv_x=\pv_y}, \m{\pb_x=\pb_y=0}; note the opposite handedness relative to the \m{\vec{\pb}}-polarised mode) and is unstable for relatively larger values of \m{K}. It may seem, at first glance, that this is simply the hydrodynamic $\alpha$-effect, also known as the H$\alpha$-effect \citep{ref:moiseev83}; however, it has been argued that the latter vanishes in incompressible turbulence \citep{ref:khomenko91}. The incompressible analogue is known as the anisotropic kinetic alpha (AKA) instability \citep{ref:frisch87}; however, as the name might suggest, this effect requires anisotropy, while our \Fig{fig:alphareference} is for an incompressible, isotropic turbulent background. Therefore, this mode remains to be interpreted. Since the focus of this work is on the turbulent-dynamo problem, we will not discuss this hydrodynamic mode any further beyond mentioning that it is stabilised by magnetic fluctuations (vanishing entirely for values of \m{m\gtrapprox 0.2}). Instead, we invite future investigations that may resolve this mystery.

\subsection{Dynamo driven by flow--current alignment}
\begin{figure}
 \centering
 \includegraphics[width=.9\textwidth]{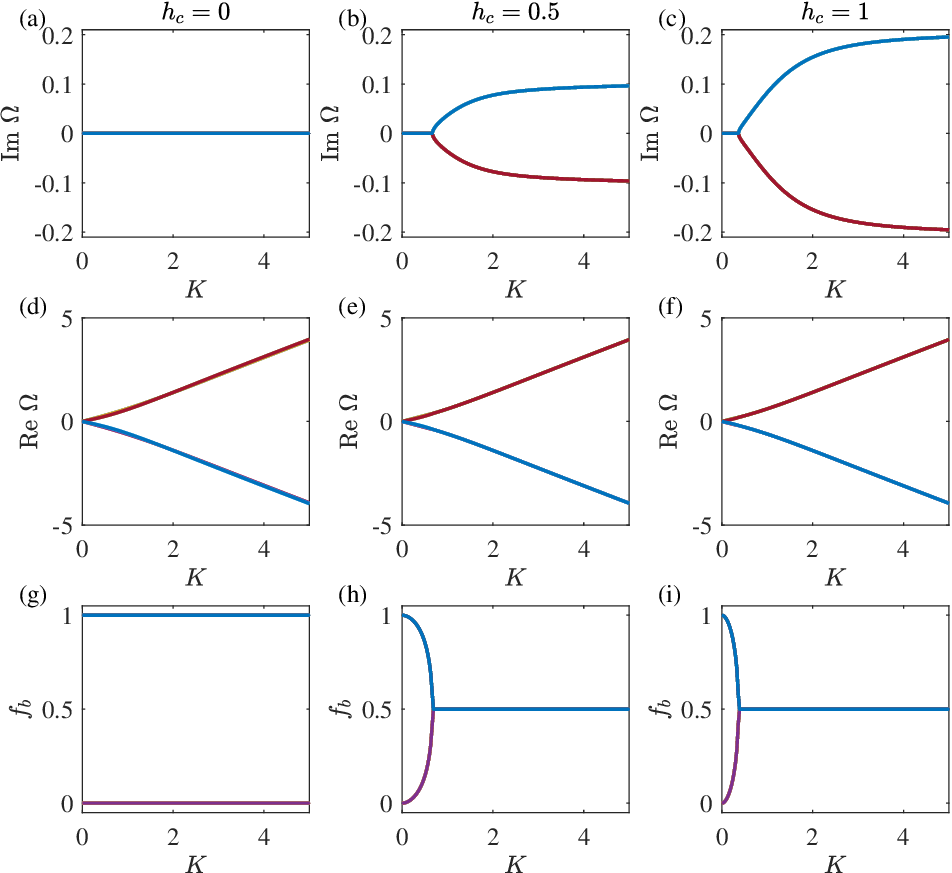}
 \caption{\label{fig:hcalone}
  The imaginary ((a)-(c)) and real ((d)-(f)) parts of modulational frequency \m{\Omega} normalised to \m{\tau^{-1}} at \m{m=1}, \m{a=h_b=h_v=0} and \m{\St=10^5} for \m{h_c=0} ((a), (d)); \m{h_c=0.5} ((b), (e)); and \m{h_c=1} ((c), (f)). The magnetic energy fraction \m{f_b\doteq \pb^2/(\pv^2+\pb^2)} of the corresponding eigenmodes is presented in panels (g)-(i). The frequencies are given in units of the inverse turnover time \m{\tau^{-1}} and wavevectors are given in units of the inverse characteristic eddy size \m{l^{-1}}. 
 }
 \end{figure}

Let us now examine the effect of flow--current alignment by solving \Eq{eq:subdisprel} for nonzero values of the parameter \m{h_c}. As discussed in \Sec{sec:nonlocalemf}, flow--current alignment (equivalently, an imbalance in the helicities of the Els\"asser fields) couples the mean momentum and induction equations, and the fundamental modulational dynamics associated with this property have not yet been established. Let us therefore first isolate the influence of flow--current alignment by removing all other potential effects (\m{a=h_v=h_b=0}). The resulting modes are shown in \Fig{fig:hcalone}.

It can be seen that the flow--current alignment \m{h_c} indeed drives a strange and novel dynamo effect, the properties of which we shall discuss at length in the remainder of this section. Let us first highlight that this effect is emphatically not one that can be captured under the assumption of scale separation, as the mode is stable for \m{lK\lessapprox 1}. Moreover, in the absence of viscosity or resistivity, the growth rate asymptotes to a constant nonzero value as \m{K\to\infty}, although we will show in \App{app:corrwaves} that this property is particular to the case \m{a=0}. 

In contrast to the stationary \m{\alpha^2}-dynamo, the unstable modes driven by flow--current alignment propagate as they grow. The propagation speed of these `correlation waves' is given by 
\begin{equation}
v_{\rm{ph}}^\pm=\sqrt{\flav{\fl{\vec{z}}^\mp\cdot\fl{\vec{z}}^\mp}/3}
\end{equation}
for \m{\av{\vec{z}}^\pm} polarised modes, and they are further discussed in \App{app:corrwaves}. Furthermore, for \m{a=0} these modes are polarised such that the associated energy is split evenly between the mean flow and magnetic fields, or equivalently, between \m{\av{\vec{z}}^+} and \m{\av{\vec{z}}^-}. Note that this means that a \m{\flav{\fl{\vec{v}}\cdot\fl{\vec{j}}}}-dynamo drives a net conversion of kinetic energy to magnetic energy as long as the turbulent magnetic energy is below equipartition (\m{m<1}). As will be discussed shortly, the energy balance for the \m{a\neq0} case is more complicated, but the ultimate consequence of dynamo action is the same.

\subsubsection{Interaction with kinetic helicity}
\begin{figure}
 \centering
 \includegraphics[width=.85\textwidth]{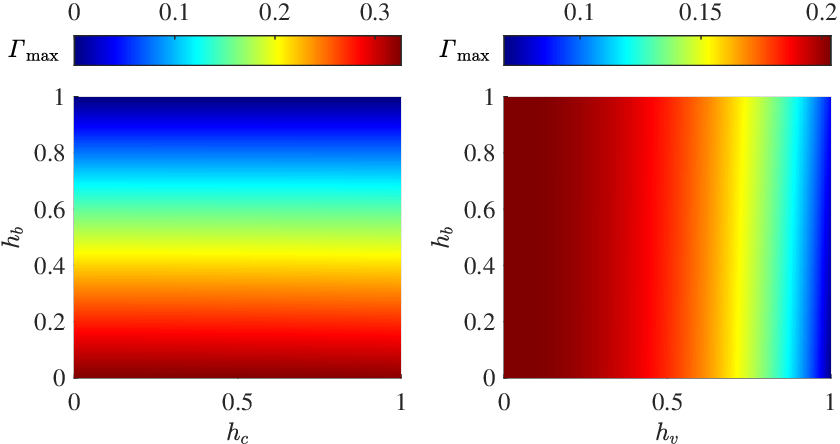}
 \caption{\label{fig:dynamocomp}
The maximum growth rate \m{\Gamma_{\mathrm{max}}} (colourbars) of the: (a)  \m{\alpha^2}-dynamo vs. flow--current alignment (\m{h_c}) and current helicity (\m{h_b}); (b) \m{\flav{\fl{\vec{v}}\cdot\fl{\vec{j}}}}-dynamo vs. the dimensionless kinetic (\m{h_v}) and current (\m{h_b}) helicities. The growth rates are given in units of the inverse turnover time \m{\tau^{-1}} and wavevectors are given in units of the inverse characteristic eddy size \m{l^{-1}}.
 }
 \end{figure}

Although the EMF for isotropic MHD turbulence can be neatly split into the sum of separate `effects' (for example, the term proportional to the mean flow that arises from \m{\flav{\fl{\vec{v}}\cdot\fl{\vec{j}}}}), these additive contributions to the EMF certainly do not guarantee additive contributions to the solutions of \Eq{eq:subdisprel}, and the various statistical properties of the turbulence can interact in non-obvious ways. To illustrate this, \Fig{fig:dynamocomp} shows the dependence of the maximum growth rate of the \m{\flav{\fl{\vec{v}}\cdot\fl{\vec{j}}}}-dynamo on the kinetic and current helicities. This figure also compares said maximum growth rate with that of the \m{\alpha^2}-dynamo, particularly in how it depends on flow--current alignment and the current helicity.

It can be seen that although flow--current alignment has almost no effect on the \m{\alpha^2}-dynamo (as opposed to current helicity, which can exactly cancel the destabilizing drive for \m{h_v=mh_b}), kinetic helicity partially suppresses the \m{\flav{\fl{\vec{v}}\cdot\fl{\vec{j}}}}-dynamo. Interestingly, the \m{\flav{\fl{\vec{v}}\cdot\fl{\vec{j}}}}-dynamo is not perceptibly effected by current helicity. Let us therefore, in the remainder of this section, take \m{h_v=mh_b}, so that the \m{\alpha^2}-dynamo is completely stabilised. This will allow us to focus on the \m{\flav{\fl{\vec{v}}\cdot\fl{\vec{j}}}}-dynamo and pin down the dimension of the parameter space that is irrelevant for the dynamics of interest.

\begin{figure}
 \centering
 \includegraphics[width=.9\textwidth]{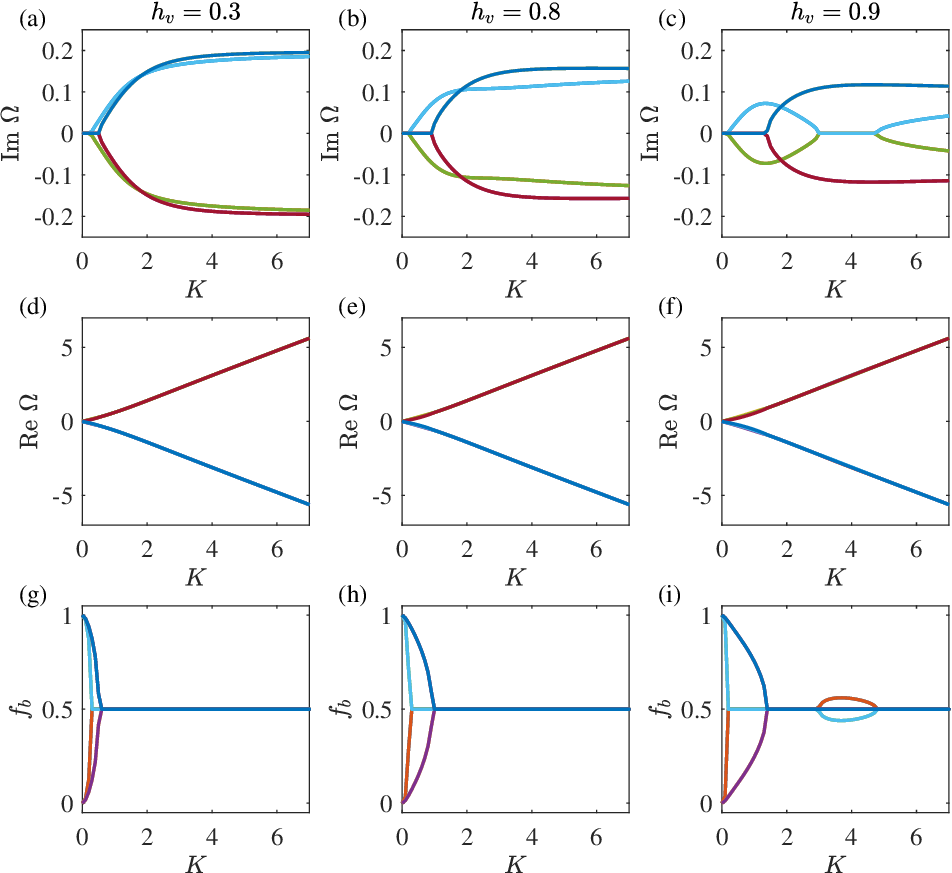}
 \caption{\label{fig:hctogether}
   The imaginary ((a)-(c)) and real ((d)-(f)) parts of modulational frequency \m{\Omega} at \m{m=h_c=1}, \m{a=0} and \m{\St=10^5} for \m{h_v=h_b=0.3} ((a), (d)); \m{h_v=h_b=0.8} ((b), (e)); and \m{h_v=h_b=0.9} ((c), (f)). The magnetic energy fraction \m{f_b\doteq \pb^2/(\pv^2+\pb^2)} of the corresponding eigenmodes are given in panels (g)-(i). The frequencies are given in units of the inverse turnover time \m{\tau^{-1}} and wavevectors are given in units of the inverse characteristic eddy size \m{l^{-1}}.
 }
 \end{figure}

Beyond a simple suppression of the maximum growth rate, kinetic helicity can qualitatively impact the \m{\flav{\fl{\vec{v}}\cdot\fl{\vec{j}}}}-driven mode properties shown in \Fig{fig:hcalone}. As seen in \Fig{fig:hctogether}, kinetic helicity breaks the degeneracy of the modes shown in \Fig{fig:hcalone}, shifting one unstable branch towards higher \m{K} and the other towards lower \m{K}. For brevity, let us refer to the former mode as \m{M_1} and the latter as \m{M_2} (each mode label refers to two separate solutions of \Eq{eq:subdisprel} with opposite signs of Re \m{\Omega}). As shown in \Fig{fig:hctogether}, at sufficiently high values of \m{h_v}, \m{M_2} splits into two unstable \m{K} regions at low and high \m{K}. The modes are both circularly polarised in both \m{\pz^+} and \m{\pz^-}, but with opposite handedness (\m{\pz_x^\pm=\ii\pz_y^\pm} for \m{M_1} and \m{\ii\pz_x^\pm=\pz_y^\pm} for \m{M_2}).

\subsubsection{Cross-helicity interaction} 
\begin{figure}
 \centering
 \includegraphics{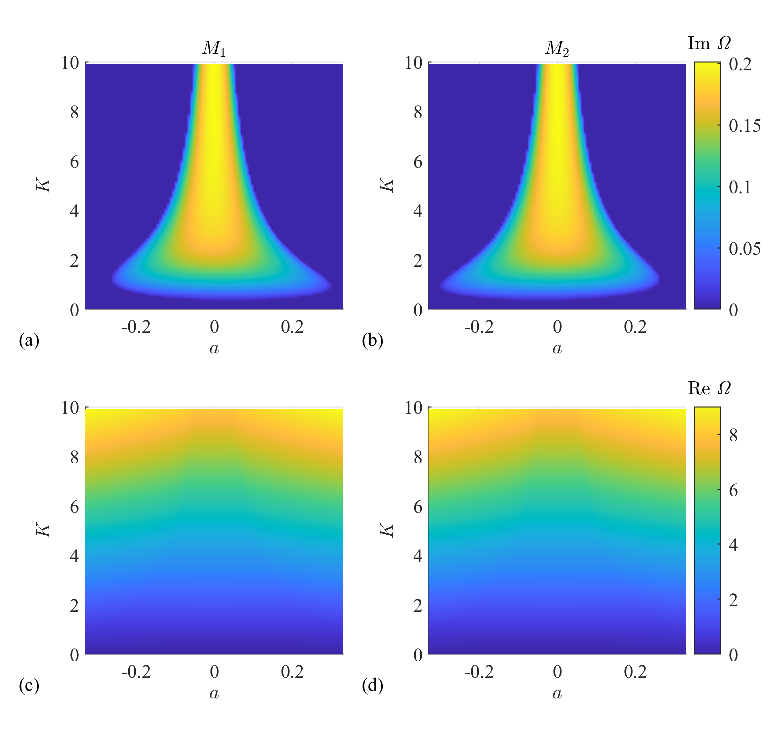}
 \caption{\label{fig:aKsurfs}
   The imaginary ((a),(b)) and real ((c),(d)) parts of modulational frequency \m{\Omega} vs. \m{K} and \m{a} at \m{m=h_v=h_b=h_c=1} and \m{\St=10^5}, for both \m{\flav{\fl{\vec{v}}\cdot\fl{\vec{j}}}}-driven modes \m{M_1} ((a), (c)) and \m{M_2} ((b), (d)). The frequencies are given in units of the inverse turnover time \m{\tau^{-1}} and wavevectors are given in units of the inverse characteristic eddy size \m{l^{-1}}.
 }
 \end{figure}

Let us now examine the impact of cross-helicity on the \m{\flav{\fl{\vec{v}}\cdot\fl{\vec{j}}}}-dynamo. As can be seen in \Fig{fig:aKsurfs}, the overall effect of cross-helicity is to suppress the \m{\flav{\fl{\vec{v}}\cdot\fl{\vec{j}}}}-dynamo and shift the instability range to smaller \m{K}, for both \m{M_1} and \m{M_2}.

 At first, this may seem like bad news for the \m{\flav{\fl{\vec{v}}\cdot\fl{\vec{j}}}}-dynamo, since cross-helicity is a measure of the imbalance between the Els\"asser fields (\m{\flav{\fl{\vec{z}}^+\cdot\fl{\vec{z}}^+}-\flav{\fl{\vec{z}}^-\cdot\fl{\vec{z}}^-}=2\flav{\fl{\vec{v}}\cdot\fl{\vec{b}}}}), and flow--current alignment is a measure of the imbalance between the Els\"asser helicities (\m{\flav{\fl{\vec{z}}^+\cdot\fl{\vec{w}}^+}-\flav{\fl{\vec{z}}^-\cdot\fl{\vec{w}}^-}=2\flav{\fl{\vec{v}}\cdot\fl{\vec{j}}}}). Although they both capture some form of imbalance between the Els\"asser fields, cross-helicity and flow--current alignment are entirely independent properties of MHD turbulence (determined by the spectra \m{E_A} and \m{H_A}, respectively). Whether such turbulence is realised in astrophysical environments or other environments of interest is a question that we leave to future work.

 \subsubsection{Dependence on \m{\St}}
\begin{figure}
 \centering
 \includegraphics{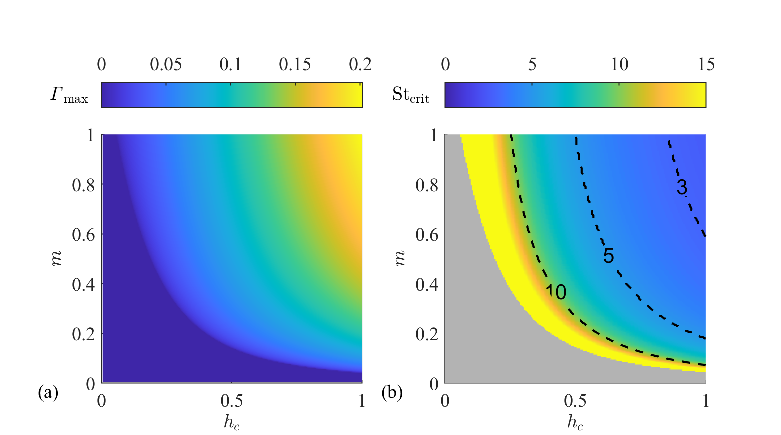}
 \caption{\label{fig:Stcrit}
(a) The maximum growth rate \m{\Gamma_{\mathrm{max}}} normalised to \m{\tau^{-1}} of \m{\flav{\fl{\vec{v}}\cdot\fl{\vec{b}}}}-dynamo, at \m{\St=10^5} and (b) critical St, \m{\St_{\mathrm{crit}}}, required for instability vs. \m{h_c} and \m{m}, both at \m{a=h_v=h_b=0}. Note that \m{\St_{\mathrm{crit}}} can only be defined for parameter values where the \m{\flav{\fl{\vec{v}}\cdot\fl{\vec{b}}}}-dynamo is unstable; gray region on (b) indicates values of \m{(m,h_c)} for which there is no instability at any value of \m{\St}. The colourbar is limited to the values \m{(0,15)} for visibility, although \m{\St_{\mathrm{crit}}} reaches much larger values near the stability boundary. 
 }
 \end{figure}

 The results presented in this section thus far are truly quasilinear, in that they are obtained in the \m{\St\to \infty} limit of \Eq{eq:modWME}. For `realistic' large-\m{\Rm} turbulence, \m{\St} is expected to be order-one. Let us therefore explore how strongly damped the \m{\flav{\fl{\vec{v}}\cdot\fl{\vec{j}}}}-dynamo is by eddy--eddy collisions, at least within the simple MTA-type model used in \Eq{eq:modWME}. Figure~\ref{fig:Stcrit} panel (a) shows the maximum growth rate of the \m{\flav{\fl{\vec{v}}\cdot\fl{\vec{j}}}}-dynamo over the \m{(m,h_c)} parameter space, and panel (b) shows the critical \m{\St} below which the fastest growing \m{\flav{\fl{\vec{v}}\cdot\fl{\vec{j}}}}-mode is completely stabilised. It can be seen that although the \m{\flav{\fl{\vec{v}}\cdot\fl{\vec{j}}}}-dynamo can still be unstable at \m{\St\sim 1}, it is, in general, far less robust to the damping effect of decorrelations than the \m{\alpha^2}-dynamo, whose growth rate scales linearly with \m{\tau_c} in the \m{\St\to 0} limit.
 \subsubsection{Discussion}
 
 In summary, we have shown that correlations between the turbulent flow and current, \m{\flav{\fl{\vec{v}}\cdot\fl{\vec{j}}}}, support a previously unknown mechanism of mean-field dynamo.
The curious properties of this dynamo warrant further investigation in a number of directions that we leave to future research.

Firstly, as we have shown, the \m{\flav{\fl{\vec{v}}\cdot\fl{\vec{j}}}}-dynamo is notably less robust to the effect of eddy--eddy interactions than the classic kinetic-helicity-driven \m{\alpha}-effect, at least within the simple MTA-type closure \Eq{eq:modWME}. Targeted DNS with the Els\"asser fields helically driven with opposite handedness could clarify if this effects persists beyond this simple closure. There is then the separate question of whether such an imbalance in Els\"asser helicities naturally occurs in astrophysical systems, \eg by the action of some instability. The relevance and broader significance of the \m{\flav{\fl{\vec{v}}\cdot\fl{\vec{j}}}}-dynamo, beyond being a curious collective effect that emerges in the quasilinear limit, hinges on such further investigations.

Although whether it is anything more remains to be seen, the \m{\flav{\fl{\vec{v}}\cdot\fl{\vec{j}}}}-dynamo is certainly an interesting and previously unrecognised collective effect of isotropic MHD turbulence, and it is worth 
pondering its properties further. Perhaps most notably, although it is a mean-field dynamo in the sense that it is a mode of the mean-field system, it is clearly not a large-scale dynamo. For the ideal balanced case shown in 
\Fig{fig:hcalone}, the \m{\flav{\fl{\vec{v}}\cdot\fl{\vec{j}}}}-modes do not have a \m{K} at which the growth rate is maximised, instead asymptoting to its maximum growth rate as \m{K\to\infty}. (This, of course, would be cut off by dissipation in practice.)
What kind of physical phenomenon does such a solution describe? As the growth rate is mostly insensitive to \m{K} at large \m{K}, perhaps this describes the system's propensity to form coherent structures, as any initial perturbation 
would roughly retain its shape while getting steeper (due to the effective high-pass filter) and growing exponentially in the linear stage. Perhaps this is some signature of the small-scale dynamo that survives the quasilinear treatment. 
We leave this as an open question for future work, but highlight the stabilizing effect of cross-helicity as a potential clue that may lead to the desired physical insight if pursued. 

\section{Summary}\label{sec:summary}
This paper aims to advance the existing understanding of self-organization in MHD turbulence by studying mean-field formation as a modulational instability of the underlying turbulence. 

In the first part of the paper (sections \ref{sec:mfwkderiv} and \ref{sec:linmod}), we propose a wave-kinetics-based extension to mean-field theory, which we term mean-field wave kinetics (MFWK). MFWK is different from previous mean-field approaches in that: (a) it does not assume scale separation between the mean fields and turbulence, and (b) it self-consistently applies the mean-field treatment to the full MHD equations rather than the induction equation alone. We also introduce the modulational-instability formulation of mean-field formation, in which the equations of MFWK are linearised around statistically homogeneous turbulent equilibria.

In the second part of the paper (sections \ref{sec:nonlocalemf} and \ref{sec:modmodes}), we apply the MFWK formalism to obtain two main results pertaining to the turbulent dynamo. 

The first of these results is an analytical expression for the nonlocal-response kernel relating the turbulent electromotive force to the mean magnetic and velocity fields for generic MHD turbulence. To the best of our knowledge, this is the first time such an expression has been derived, as opposed to inferred through direct numerical simulations (DNS). Our result is in agreement with DNS and previous analytical findings in the relevant limits, with the important exception of the dependence of the EMF on the mean flow. The disagreement between our calculation of this effect and the previous result \citep{ref:radler10} highlights the importance of a self-consistent treatment of the flow. 

The second main result is our prediction of a novel mean-field dynamo effect that is driven by correlations between the turbulent flow and current, \m{\flav{\fl{\vec{v}}\cdot\fl{\vec{j}}}}. We emphasise that such an effect cannot be captured with the usual kinematic and scale-separated approach. Although the \m{\flav{\fl{\vec{v}}\cdot\fl{\vec{j}}}}-dynamo is generally less robust than the well-known \m{\alpha^2}-dynamo, it has the important distinction that it is not quenched by current helicity, and is, in fact, enabled rather than suppressed by magnetic fluctuations. Whether this effect survives beyond the QLA, and whether the flow--current correlations required for its onset are relevant to astrophysical environments or other environments of interest remains to be seen. 
\section*{Acknowledgments}
SJ thanks Amitava Bhattacharjee for helpful comments.
\section*{Funding}
This research was supported by the U.S.\ Department of Energy through contract No.\ DE-AC02-09CH11466. 
\section*{Declaration of interests}
The authors report no conflict of interest.
\appendix

\section{Review of the Wigner--Weyl transform\label{app:wignerweyl}}

Here we provide a minimal, low-brow review of the Wigner--Weyl transform and list some useful properties and identities that are used throughout this work. For a more in-depth discussion, see, for example, \citet{book:tracy14} or \citet{ref:case08}.

We will write our formulae for the specific case of operators \m{\op{A}} acting on the Hilbert space of functions \m{f(\vec{x})} defined on a three-dimensional configuration space \m{\vec{x}\doteq(x_1,x_2,x_3)}. The formalism is, of course, more general, but here we prioritise ease of application to the specific problems considered in this paper over generality. We will also limit our consideration to scalar operators and scalar functions, as matrix-valued operators and functions can be treated element-wise with the scalar formulae.

The Wigner--Weyl transform maps an operator \m{\op{A}} onto a function \m{A} on the \m{2\times3}-dimensional phase space \m{(\vec{x},\vec{k})}:

\begin{equation}
\begin{aligned}
        A(\vec{x},\vec{k})&=\int \dd\vec{s}\,\ee^{-\ii\vec{k}\cdot\vec{s}}\braket{\vec{x}+\vec{s}/2|\op{A}|\vec{x}-\vec{s}/2}.\\
\end{aligned}
\end{equation}
Here, the kets \m{\ket{\vec{x}}} are the eigenstates of the position operator \m{\op{\vec{x}}}, normalised such that \m{\braket{\vec{x}'|\op{\vec{x}}|\vec{x}}=\vec{x}\delta(\vec{x}-\vec{x}')}.
This projection is known as the the Weyl image or symbol of \m{\op{A}}.

The inverse Wigner--Weyl transform is defined:
\begin{equation}
    \op{A}=\frac{1}{(2\pi)^3}\int\dd\vec{x}\int\dd\vec{k} \int \dd \vec{s} \,\ee^{-\ii\vec{k}\cdot\vec{s}}A(\vec{x},\vec{k})\ket{\vec{x}-\vec{s}/2}\bra{\vec{x}+\vec{s}/2}.
\end{equation}
It follows that the matrix elements of \m{\op{A}} in the coordinate representation, \m{\mc{A}(\vec{x},\vec{x}')\doteq} 
 \m{\braket{\vec{x}|\op{A}|\vec{x}'}}, are connected with \m{A} via
\begin{equation}
    \mc{A}(\vec{x},\vec{x}')=\frac{1}{(2\pi)^3}\int\dd\vec{k}\,\ee^{-\ii\vec{k}\cdot(\vec{x}-\vec{x}')}A\Big(\frac{\vec{x}+\vec{x}'}{2},\vec{k}\Big), 
\end{equation}
and in particular,
\begin{equation}\label{eq:useful}
    \mc{A}(\vec{x},\vec{x})=\frac{1}{(2\pi)^3}\int\dd\vec{k}\,A(\vec{x},\vec{k}). 
\end{equation}

The identity, position, and momentum operators have intuitive mappings:
\begin{equation}
    \op{1}\Leftrightarrow1, \qquad \op{\vec{x}}\Leftrightarrow \vec{x}, \qquad \op{\vec{k}}\Leftrightarrow \vec{k},
\end{equation}
where we use \m{\Leftrightarrow} to denote the Wigner--Weyl correspondence between operators and their symbols. Also, pure functions of the position and momentum operators map on to the same functions of the corresponding phase-space coordinate:
\begin{equation}
    f(\op{\vec{x}})\Leftrightarrow f(\vec{x}), \qquad g(\op{\vec{k}})\Leftrightarrow g(\vec{k})
\end{equation}
for any functions \m{f} and \m{g}.

More generally, the product of operators \m{\op{C}=\op{A}\op{B}} maps in the following way:
\begin{equation}\label{eq:opproduct}
    \op{A}\op{B}\Leftrightarrow A(\vec{x},\vec{k})\star B(\vec{x},\vec{k}),
\end{equation}
where \m{\star} is the Moyal star product. This product is defined as
\begin{equation}\label{eq:moyalstar}
    A(\vec{x},\vec{k})\star B(\vec{x},\vec{k})\doteq A(\vec{x},\vec{k})\ee^{\ii\op{\mc{L}}/2}B(\vec{x},\vec{k}),
\end{equation}
where \m{\mc{L}} is the Janus operator, 
\begin{equation}\label{eq:janusop}
    \op{\mc{L}}\doteq \overleftarrow{\pd_{\vec{x}}}\cdot\overrightarrow{\pd_{\vec{k}}}-\overleftarrow{\pd_{\vec{k}}}\cdot\overrightarrow{\pd_{\vec{x}}},
\end{equation}
and the arrows indicate the directions in which the derivatives act, so that \m{A\op{\mc{L}}B=\{A,B\}}, where \m{\{...\,,\,...\}} is the canonical Poisson bracket,
\begin{equation}
    \{A,B\}\doteq(\pd_{\vec{x}}A)\cdot(\pd_{\vec{k}}B)-(\pd_{\vec{k}}A)\cdot(\pd_{\vec{x}}B).
\end{equation}
In the main part of the paper, we make frequent use of the following identities:
\begin{equation}\label{eq:starshift}
A(\vec{k})\star\ee^{\ii\vec{K}\cdot\vec{x}}=A(\vec{k}+\vec{K}/2)\ee^{\ii\vec{K}\cdot\vec{x}},\qquad \ee^{\ii\vec{K}\cdot\vec{x}}\star A(\vec{k})=A(\vec{k}-\vec{K}/2)\ee^{\ii\vec{K}\cdot\vec{x}},
\end{equation}
where \m{\vec{K}} is a constant. For other potentially useful properties of the Moyal star, see, for example, \citet{book:tracy14}.
\section{Quadratic invariants of mean-field wave kinetics\label{app:conservationlaws}}
In this appendix, we show that the quasilinear MFWK equations \Eq{eq:summary} derived in \Sec{sec:mfwkderiv} conserve two key invariants of the original MHD model \Eq{eq:mhd}: energy and cross-helicity. In what follows, it will be useful to work with the Fourier transformed ideal system:
\begin{subequations}\label{eq:fouriersummary}
\begin{gather}
        \pd_t \av{\vec{w}}^\pm(\vec{q}) = \int\frac{\dd\vec{q}'}{(2\pi)^3}[\vec{q}\cdot\av{\vec{z}}^\pm(\vec{q}')][\vec{q}\times\av{\vec{z}}^\pm(\vec{q}-\vec{q}')]+\vec{S}^\pm(\vec{q}),\\
\begin{aligned}
    \ii\pd_t\matr{W}(\vec{q},\vec{k})=\int\frac{\dd\vec{q}'}{(2\pi)^3}\Bigg\{&\matr{H}\Big(\vec{q}',\vec{k}+\frac{\vec{q}-\vec{q}'}{2}\Big)\matr{W}\Big(\vec{q}-\vec{q}',\vec{k}-\frac{\vec{q}'}{2}\Big)\\
    &-\matr{W}\Big(\vec{q}-\vec{q}',\vec{k}+\frac{\vec{q}'}{2}\Big)\matr{H}^\dagger\Big(\vec{q}',\vec{k}-\frac{\vec{q}-\vec{q}'}{2}\Big)\Bigg\},
\end{aligned}\\
S_i^\pm(\vec{q})=\epsilon_{ijk}\int\frac{\dd\vec{k}}{(2\pi)^3}q_j\Big(k_l+\frac{q_l}{2}\Big)W_{kl}^{\pm\mp}(\vec{q},\vec{k}),\\
        \begin{gathered}
H^{\pm\pm}_{ij}(\vec{q},\vec{k})=[\vec{k}\cdot\vec{\av{z}}^\mp(\vec{q})]\Big(\delta_{ij}-\frac{(k_i+\frac{q_i}{2})q_j}{(\vec{k}+\frac{\vec{q}}{2})^2}\Big),\\
        H_{ij}^{\pm\mp}(\vec{q},\vec{k})=q_j\av{z}_i^\pm(\vec{q})-[\vec{k}\cdot\vec{\av{z}}^\pm(\vec{q})]\frac{(k_i+\frac{q_i}{2})q_j}{(\vec{k}+\frac{\vec{q}}{2})^2}.
        \end{gathered}
\end{gather}
\end{subequations}
Note also that incompressibility gives
\begin{equation}
    \Big(k_l+\frac{q_l}{2}\Big)W^{\sigma_1\sigma_2}_{lm}(\vec{q},\vec{k})=    \Big(k_m-\frac{q_m}{2}\Big)W^{\sigma_1\sigma_2}_{lm}(\vec{q},\vec{k})=0.
\end{equation}
The simultaneous conservation of energy and cross-helicity implies that
\begin{equation}
    \frac{\dd E^{\pm}}{\dd t}=0,
\end{equation}
where 
\begin{equation}
    E^{\pm}=\av{E}^\pm+\fl{E}^\pm=\int\dd \vec{x}\Big(|\vec{\av{z}}^\pm|^2+|\vec{\fl{z}}^\pm|^2\Big).
\end{equation}

Let's start with the mean field energies:
\begin{equation}
\begin{aligned}
    \frac{\dd \av{E}^\pm}{\dd t}&=\frac{\dd}{\dd t}\int\dd\vec{x}\,|\vec{\av{z}}^\pm|^2\\
&=2\int\dd\vec{x}\,\vec{\av{z}}\cdot\pd_t\vec{\av{z}}\\
&=2\int\dd\vec{x}\,\vec{\av{z}}^\pm\cdot[-(\vec{\av{z}}^\mp\cdot\nabla)\vec{\av{z}}^\pm-\nabla\av{P}-\flav{(\vec{\fl{z}}^\mp\cdot\nabla)\vec{\fl{z}}^\pm}-\nabla\flav{\fl{P}}]\\
&=2\int\dd\vec{x}\Bigg\{\nabla\cdot\Big(\frac{1}{2}|\vec{\av{z}}^\pm|^2\vec{\av{z}}^\mp-(\av{P}+\fl{P)}\vec{\av{z}}^\pm\Big) -\ii\av{z}^\pm_l\int\frac{\dd\vec{k}}{(2\pi)^3}\,k_m\star W^{\pm\mp}_{lm}\Bigg\}\\
&=-2\ii\int\frac{\dd\vec{k}}{(2\pi)^3}\int\frac{\dd\vec{q}}{(2\pi)^3}\,\av{z}_l^\pm(-\vec{q})\Big(k_m+\frac{q_m}{2}\Big)W_{lm}^{\pm\mp}(\vec{q},\vec{k})\\
&=2\im\int\frac{\dd\vec{k}}{(2\pi)^3}\int\frac{\dd\vec{q}}{(2\pi)^3}\,\av{z}_l^\pm(-\vec{q}) W^{\pm\mp}_{lm}(\vec{q},\vec{k})q_m.
\end{aligned}
\end{equation}
Now for the energies in the turbulent fields:
\begin{equation}
\begin{aligned}
    \frac{\dd \fl{E}^\pm}{\dd t}&=\frac{\dd}{\dd t}\int\dd\vec{x}\,|\vec{\fl{z}}^\pm(\vec{x})|^2\\
&=\int\dd\vec{x}\int\frac{\dd\vec{k}}{(2\pi)^3}\,\pd_tW^{\pm\pm}_{ll}(\vec{x},\vec{k})\\
   &=2\im\int \dd\vec{x}\int\frac{\dd\vec{k}}{(2\pi)^3}\Big[H^{\pm\pm}_{lm}\star W^{\pm\pm}_{ml}+H^{\pm\mp}_{lm}\star W^{\mp\pm}_{ml}\Big]\\
      &=2\im\int \frac{\dd\vec{k}}{(2\pi)^3}\int\frac{\dd\vec{q}}{(2\pi)^3}\Big[\,H^{\pm\pm}_{lm}(-\vec{q},\vec{k}) W^{\pm\pm}_{ml}(\vec{q},\vec{k})\\&\qquad\qquad\qquad\qquad\qquad\,\,\,+H^{\pm\mp}_{lm}(-\vec{q},\vec{k}) W^{\mp\pm}_{ml}(\vec{q},\vec{k})\Big]\\
            &=2\im\int \frac{\dd\vec{k}}{(2\pi)^3}\int\frac{\dd\vec{q}}{(2\pi)^3}\Bigg\{\Bigg[[\vec{k}\cdot\vec{\av{z}}^\mp(-\vec{q})]\Big(\delta_{lm}+\frac{(k_l-\frac{q_l}{2})q_m}{(\vec{k}-\frac{\vec{q}}{2})^2}\Big)\Bigg] W^{\pm\pm}_{ml}(\vec{q},\vec{k})\\
            &\qquad\qquad\qquad\qquad\;+\Bigg[-q_m\av{z}_l^\pm(-\vec{q})-[\vec{k}\cdot\vec{\av{z}}^\pm(\vec{q})]\frac{(k_l-\frac{q_l}{2})q_m}{(\vec{k}+\frac{\vec{q}}{2})^2}\Bigg] W^{\mp\pm}_{ml}(\vec{q},\vec{k})\Bigg\}\\
            &=-2\im\int \frac{\dd\vec{k}}{(2\pi)^3}\int\frac{\dd\vec{q}}{(2\pi)^3}\,\av{z}_l^\pm(-\vec{q})W_{lm}^{\pm\mp}(\vec{q},\vec{k})q_m.
\end{aligned}
\end{equation}
Therefore, \m{\av{E}^\pm+\fl{E}^\pm=\const}, and energy and cross-helicity are conserved for our quasilinear system.
\section{diagonalising transformation for 2-D MHD\label{app:2d}}
In this appendix, we show that the Hamiltonian for the fluctuating fields in 2-D is approximately diagonalisable in the strong magnetic field limit, \m{|\av{\vec{b}}|\gg|\av{\vec{v}}|}.

Consider 2-D dynamics in which \m{\vec{z}^\pm} lie in the \m{(x, y)} plane and \m{\pd_z = 0}. In this case, the only potentially nonzero component of \m{\vec{w}^\pm} is the \m{z}-component, \m{w^\pm \doteq (\nabla\times\vec{z}^\pm)_z}. Then, the vector equation \eq{eq:vecw} can be replaced with a scalar equation for \m{w^\pm}: 
\begin{gather}
\label{eq:scalarw}
   \pd_t w^\pm=-(\vec{z}^\mp\cdot\nabla)w^\pm+\nabla \vec{z}^\mp : \nabla\nabla \op{k}^{-2}w^\pm,
\end{gather}
Splitting the Els\"asser vorticities in fluctuating and averaged parts ($w^\pm=\av{w}^\pm+\fl{w}^\pm$ with $\flav{\fl{w}}=0$), one obtains the following equation for the fluctuating fields \m{\fl{w}^\pm}:
    \begin{equation}\label{eq:wfluctuation}
     \begin{aligned}
      \partial_t \fl{w}^\pm=&(-(\av{\vec{z}}^\mp\cdot\nabla)+\nabla \av{\vec{z}}^\mp : \nabla\nabla\op{k}^{-2} )\fl{w}^\pm+[-(\nabla^2\av{\vec{z}}^\pm\cdot\nabla)-\nabla \av{\vec{z}}^\pm:\nabla\nabla \op{k}^{-2}]\fl{w}^\mp,
 \end{aligned} 
 \end{equation}
where \m{\av{\vec{z}}^\pm\doteq \ii\op{k}^{-2}(\op{k}_y \av{w}^\pm,-\op{k}_x \av{w}^\pm, 0)^\intercal} and the terms nonlinear in the fluctuating fields have been neglected.

Equation (\ref{eq:wfluctuation}) can be written as a vector Schr\"odinger equation for the field variables \m{\xi^\pm=\op{k}^{-1}\fl{w}^\pm}:
\begin{equation}\label{eq:vecschro}
    \ii\pd_t\vec{\xi}=\op{\matr{\mc{H}}}\vec{\xi},
\end{equation}
where \m{\vec{\xi}\doteq(\xi^+,\xi^-)^\intercal} and the Hamiltonian \m{\op{\matr{\mc{H}}}} is given by 
\begin{equation}
    \op{\matr{\mc{H}}}\doteq\begin{pmatrix}
         \op{\mc{H}}^{+}&\op{\chi}^+\\
         \op{\chi}^-& \op{\mc{H}}^{-}
    \end{pmatrix},
\end{equation}
where we have defined
 \begin{gather}\label{eq:hdiag}
    \begin{aligned}
    \op{\mc{H}}^{\pm}&\doteq\op{k}^{-1}[(\av{\vec{z}}^\pm\cdot\op{\vec{k}})-i((\nabla\av{\vec{z}}^\pm\cdot\op{\vec{k}})\cdot\op{\vec{k}} )\op{k}^{-2}]\op{k}\\
    &=\op{k}^{-1}\op{k}_l(\av{z}^\mp\cdot\op{\vec{k}})\op{k}_l\op{k}^{-1},
    \end{aligned}
    \end{gather}
and 

    \begin{gather}\label{eq:hodiag}
    \begin{aligned}
    \op{\chi}^\pm&\doteq \op{k}^{-1}[i((\nabla\av{\vec{z}}^\pm\cdot\op{\vec{k}})\cdot\op{\vec{k}} )\op{k}^{-2}+(\nabla^2\av{\vec{z}}^\pm\cdot\op{\vec{k}})\op{k}^{-2}]\op{k}\\
    &=\ii\op{k}^{-1}\op{k}_l(\pd_l\av{z}^\mp\cdot\op{\vec{k}})\op{k}^{-1}.
    \end{aligned}
    \end{gather}
Note that the operators $\op{\mc{H}}^{\pm}$ are exactly Hermitian. 

Suppose we are interested in the quasiparticle dynamics up to \m{\mc{O}(\epsilon^2)} in the GO parameter \m{\epsilon\doteq l/L}. Since the off-diagonal elements of the Hamiltonian \m{\op{\chi}^{\pm}} are \m{\mc{O}(\epsilon)}, one cannot treat the Els\"asser vorticities \m{\fl{w}^\pm} as decoupled scalar waves. Let us therefore consider the transformed variable \m{\vec{\psi}\doteq\op{\matr{U}}\vec{\xi}}, where the operator \m{\op{\matr
U}} is taken to be of the form 
\begin{equation}\label{eq:transformation}
    \op{\matr{U}}=\begin{pmatrix}
        1&\op{a}\\
        \op{b}& 1
    \end{pmatrix}, 
\end{equation}
which has the inverse
\begin{equation}
        \op{\matr{U}}^{-1}=\begin{pmatrix}
               \op{g}_1&-\op{g}_1\op{a}\\
        -\op{g}_2\op{b}& \op{g}_2
    \end{pmatrix},
\end{equation}
where 
\begin{equation}
    \op{g}^1\doteq(1-\op{a}\op{b})^{-1}, \qquad\op{g}_2\doteq(1-\op{b}\op{a})^{-1}.
\end{equation}
If \m{\pd_t\op{\matr{U}}} is negligible, \Eq{eq:vecschro} can be rewritten as 
\begin{equation}
    \ii\pd_t\vec{\psi}=\op{\matr{\mathrm{H}}}\vec{\psi}
\end{equation}
with the transformed Hamiltonian 
\begin{equation}
\begin{aligned}
        \op{\matr{\mathrm{H}}}&\doteq\op{\matr{U}}\op{\matr{\mc{H}}}\op{\matr{U}}^{-1}\\
        &=\begin{pmatrix}
        \op{g}_1(\op{\mc{H}}^++\op{\chi}^+\op{b}-\op{a}\op{\chi}^--\op{a}\op{\mc{H}}^-\op{b})    &\op{g}_1(\op{\mc{H}}^+\op{a}+\op{\chi}^+-\op{a}\op{\chi}^-\op{a}-\op{a}\op{\mc{H}}^-)\\
        \op{g}_2(-\op{b}\op{\mc{H}}^+-\op{b}\op{\chi}^+\op{b}+\op{\chi}^-+\op{\mc{H}}^-\op{b})&\op{g}_2(-\op{b}\op{\mc{H}}^+\op{a}-\op{b}\op{\chi}^++\op{\chi}^-\op{a}+\op{\mc{H}}^-)\end{pmatrix}.
\end{aligned}
\end{equation}
To diagonalise \m{\op{\matr{\mathrm{H}}}}, the operators \m{\op{a}} and \m{\op{b}} must satisfy
\begin{subequations}\label{eq:diagcond}
    \begin{gather}
        \op{\mc{H}}^+\op{a}+\op{\chi}^+-\op{a}\op{\chi}^-\op{a}-\op{a}\op{\mc{H}}^-=0,\\
        -\op{b}\op{\mc{H}}^+-\op{b}\op{\chi}^+\op{b}+\op{\chi}^-+\op{\mc{H}}^-\op{b}=0,
    \end{gather}
\end{subequations}
which also simplifies the diagonal elements:
\begin{subequations}
    \begin{gather}
    \begin{aligned}
                \op{\mathrm{H}}_{11}&=\op{g}_1[\op{\mc{H}}^++\op{\chi}^+\op{b}-\op{a}\op{\chi}^--\op{a}(\op{b}\op{\mc{H}}^++\op{b}\op{\chi}^+\op{b}-\op{\chi}^-)]\\
                &=\op{\mc{H}}^++\op{\chi}^+\op{b},
    \end{aligned}\\
    \begin{aligned}
                \op{\mathrm{H}}_{22}&=\op{g}_2[-\op{b}(-\op{\chi}^++\op{a}\op{\chi}^-\op{a}+\op{a}\op{\mc{H}}^-)-\op{b}\op{\chi}^++\op{\chi}^-\op{a}+\op{\mc{H}}^-]\\
                &=\op{\mc{H}}^-+\op{\chi}^-\op{a},
    \end{aligned}
    \end{gather}
\end{subequations}
such that
\begin{equation}
    \op{\matr{\mathrm{H}}}=\begin{pmatrix}
        \op{\mc{H}}^++\op{\chi}^+\op{b}&0\\
        0&\op{\mc{H}}^-+\op{\chi}^-\op{a}
    \end{pmatrix}
\end{equation}
for \m{\op{a}} and \m{\op{b}} satisfying \Eq{eq:diagcond}.

In the strong magnetic field limit, \m{|\av{\vec{v}}|/|\av{\vec{b}}|\sim\mc{O}(\epsilon)}, \Eq{eq:diagcond} can be solved approximately. To leading order in \m{\epsilon}, \Eq{eq:diagcond} become:
\begin{subequations}\label{eq:diagcondapprox}
    \begin{gather}
        \op{a}(\op{\mc{H}}^+-\op{\mc{H}}^-)+\op{\chi}^+=0,\\
        \op{b}(\op{\mc{H}}^--\op{\mc{H}}^+)+\op{\chi}^-=0,
    \end{gather}
\end{subequations}
which yield the approximate solutions
\begin{equation}
    \op{a}\approx-\frac{1}{2}\op{\chi}^+\op{\mc{H}}_b^{-1}, \qquad\op{b}\approx\frac{1}{2}\op{\chi}^-\op{\mc{H}}_b^{-1},
\end{equation}
where \m{\op{\mc{H}}_b\doteq(\op{\mc{H}}^+-\op{\mc{H}}^-)/2}.

Since \m{\op{\chi}^+} and \m{\op{\chi}^-} commute within the accuracy of the approximation, we finally have
\begin{equation}
    \op{\matr{\mathrm{H}}}=\begin{pmatrix}
        \op{\mc{H}}^+-\op{\sigma}&0\\
        0&\op{\mc{H}}^-+\op{\sigma}
    \end{pmatrix}+\mc{O}(\epsilon^2),
\end{equation}
where
\begin{equation}
    \op{\sigma}\doteq\frac{1}{2}\op{\chi}^+\op{\chi}^-\op{\mc{H}}_b^{-1}.
\end{equation}
In this approximation, \m{\psi^+} and \m{\psi^-} evolve independently and are governed by the Hamiltonians \m{\op{\mc{H}}^+-\op{\sigma}} and \m{\op{\mc{H}}^-+\op{\sigma}}, respectively.

\section{Correlation waves}\label{app:corrwaves}
As can be seen in the many plots of modulational frequency vs. wavenumber \m{\Omega(K)} throughout section \ref{sec:modmodes}, the generic modulational response to a high-\m{K} mean-field perturbation takes the form of a sound-like traveling wave, \ie a wave with linear \m{\Omega(K)}. Although
instabilities driven by the various helicities may modify or dominate over this oscillatory tendency at smaller values of \m{K}, it can be seen that the real components of all solutions of \Eq{eq:subdisprel} eventually
converge to the universal sound-like solutions as \m{K\to\infty}. Let us therefore consider the \m{K\to\infty} limit of \Eq{eq:ideallim}:

\begin{subequations}
    \begin{gather}
        M_S\to\ii\Omega+\frac{\ii}{\Omega'}\Big(-\frac{1}{3}\flav{\fl{v}^2+\fl{b}^2}K^2+\frac{1}{5}\flav{\fl{w}^2+\fl{j}^2}-\frac{1}{30}\flav{\fl{w}^2-\fl{j}^2}\Big),\\
        N_S\to\frac{\ii}{\Omega'}\Big(-\frac{1}{30}\flav{\fl{w}^2+\fl{j}^2}+\frac{1}{5}\flav{\fl{w}^2-\fl{j}^2}\Big),\\
        M_A\to\frac{\ii}{\Omega'}\Big(\frac{2}{3}\flav{\fl{\vec{v}}\cdot\fl{\vec{b}}}K^2-\frac{2}{5}\flav{\fl{\vec{w}}\cdot\fl{\vec{j}}}\Big),\\
        N_A\to-\frac{\ii}{15\Omega'}\flav{\fl{\vec{w}}\cdot\fl{\vec{j}}},\displaybreak\\
        m_S\to 0,\\
        n_S\to\frac{1}{3\Omega'}\flav{\fl{\vec{v}}\cdot\nabla\times\fl{\vec{v}}}K,\\
        n_A\to\frac{1}{3\Omega'}\flav{\fl{\vec{v}}\cdot\fl{\vec{j}}}K,
    \end{gather}
\end{subequations}

where
\begin{equation}
    \begin{gathered}
        \int_0^\infty\dd k \,k^2 E_S(k)=\frac{1}{2}\flav{\fl{w}^2+\fl{j}^2},\\
        \int_0^\infty\dd k \,k^2 E_C(k)=\frac{1}{2}\flav{\fl{w}^2-\fl{j}^2},\\
        \int_0^\infty\dd k \,k^2 E_A(k)=\flav{\fl{\vec{w}}\cdot\fl{\vec{j}}},\\
        \int_0^\infty\dd k \,k^2 H_S(k)=\flav{\fl{\vec{w}}\cdot\nabla\times\fl{\vec{w}}+\fl{\vec{j}}\cdot\nabla\times\fl{\vec{j}}},\\
        \int_0^\infty\dd k \,k^2 H_C(k)=\flav{\fl{\vec{w}}\cdot\nabla\times\fl{\vec{w}}-\fl{\vec{j}}\cdot\nabla\times\fl{\vec{j}}},\\\int_0^\infty\dd k \,k^2 H_A(k)=\flav{\fl{\vec{w}}\cdot\nabla\times\fl{\vec{j}}+\fl{\vec{j}}\cdot\nabla\times\fl{\vec{w}}},\\
        \end{gathered}
\end{equation}
and as before, \m{m_A=0}.
To the leading order in \m{K}, and taking \m{\Omega'=\Omega} for simplicity, we then have
\begin{gather}
M_S=\ii\Omega-\frac{\ii}{3\Omega'}\flav{\fl{v}^2+\fl{b}^2}K^2,\\ M_A=\frac{2\ii}{3\Omega'}\flav{\fl{\vec{v}}\cdot\fl{\vec{b}}},
\end{gather}
\begin{equation}\label{eq:leadingorderpi}
    \sub{\matr{\Pi}}\doteq\frac{\ii}{\Omega}\begin{pmatrix}
    \Big(\Omega^2-\frac{\flav{\fl{\vec{z}}^-\cdot\fl{\vec{z}}^-}}{3}K^2\Big)\matr{1}_2&\matr{0}_2\\
    \matr{0}_2&\Big(\Omega^2-\frac{\flav{\fl{\vec{z}}^+\cdot\fl{\vec{z}}^+}}{3}K^2\Big)\matr{1}_2\\
            \end{pmatrix},
\end{equation}
where \m{\matr{1}_2} is the \m{2\times2} identity matrix and \m{\matr{0}_2} is the \m{2\times2} zero matrix. It can be easily seen from \Eq{eq:leadingorderpi} that the sound-like modes take the form of \m{\av{\vec{z}}^\pm} polarised modes propagating with two different speeds, \m{(\flav{\fl{\vec{z}}^\mp\cdot\fl{\vec{z}}^\mp}/3)^{1/2}}:
\begin{equation}
\Omega^2=\frac{\flav{\fl{\vec{z}}^\pm\cdot\fl{\vec{z}}^\pm}}{3}K^2,
\end{equation}
provided that the turbulence is imbalanced, \ie \m{\flav{\fl{\vec{z}}^+\cdot\fl{\vec{z}}^+}\neq \flav{\fl{\vec{z}}^-\cdot\fl{\vec{z}}^-}}. If, however, the turbulence is balanced (\m{\flav{\fl{\vec{z}}^+\cdot\fl{\vec{z}}^+}= \flav{\fl{\vec{z}}^-\cdot\fl{\vec{z}}^-}}), and \m{\Omega^2=\flav{\fl{v}^2+\fl{b}^2}K^2/3} can eliminate all of the leading order (\m{\sim K^2}) terms in \Eq{eq:leadingorderpi} then distinguishing the two modes requires consideration of higher-order corrections. 

Letting \m{\Omega^2=\flav{\fl{v}^2+\fl{b}^2}K^2/3+\delta\Omega^2}, we have to next highest order in \m{K}:
\begin{equation}
\Omega M_S\to\ii\delta\Omega^2,\qquad \Omega n_S\to\frac{\flav{\fl{\vec{v}}\cdot\fl{\vec{w}}}}{3}K, \qquad\Omega n_A\to\frac{\flav{\fl{\vec{v}}\cdot\fl{\vec{j}}}}{3}K,
\end{equation}
\begin{equation}\label{eq:nextorder}
    \sub{\matr{\Pi}}\doteq\frac{1}{\Omega}\begin{pmatrix}
    \ii\delta\Omega^2&0&0&\frac{\flav{\fl{\vec{v}}\cdot\fl{\vec{w}}}+\flav{\fl{\vec{v}}\cdot\fl{\vec{j}}}}{3}K\\
    0&\ii\delta\Omega^2&-\frac{\flav{\fl{\vec{v}}\cdot\fl{\vec{w}}}+\flav{\fl{\vec{v}}\cdot\fl{\vec{j}}}}{3}K&0\\
    0&\frac{\flav{\fl{\vec{v}}\cdot\fl{\vec{w}}}-\flav{\fl{\vec{v}}\cdot\fl{\vec{j}}}}{3}K&\ii\delta\Omega^2&0\\-
    \frac{\flav{\fl{\vec{v}}\cdot\fl{\vec{w}}}-\flav{\fl{\vec{v}}\cdot\fl{\vec{j}}}}{3}K&0&0&\ii\delta\Omega^2\\
            \end{pmatrix},
\end{equation}
and we find from the dispersion relation that 
\begin{equation}\label{eq:omegacorrection}
    \delta\Omega^2=\pm\frac{\sqrt{\flav{\fl{\vec{v}}\cdot\fl{\vec{w}}}^2-\flav{\fl{\vec{v}}\cdot\fl{\vec{j}}}^2}}{3}K.
\end{equation}
In other words, these solutions support a finite growth rate even as \m{K\to\infty}:
\begin{equation}
    \Gamma_\infty^2=\frac{\flav{\fl{\vec{v}}\cdot\fl{\vec{j}}}^2-\flav{\fl{\vec{v}}\cdot\fl{\vec{w}}}^2}{2\flav{\fl{v}^2+\fl{b}^2}},
\end{equation}
and the polarizations are relatively complicated as can be seen from \Eq{eq:nextorder} and \Eq{eq:omegacorrection}.

Finally, note that the decorrelation term can be restored in the calculations above by setting \m{\Omega^2\to\Omega\Omega'}. Since \m{\Omega\Omega'=(\Omega+\ii\tau_c^{-1}/2)^2-(\tau_c^{-1}/2)^2}, we can see that the modes discussed above are damped at a rate \m{\tau_c^{-1}/2}.

\bibliographystyle{jpp}

\bibliography{bib-main}

\end{document}